\documentclass[preprint,11pt,times]{elsarticle}
\usepackage{natbib}
\usepackage{amsmath,amsfonts,amssymb,bbm}
\usepackage{geometry} 
\usepackage{setspace} 
\usepackage{xargs}   
\usepackage[nice]{nicefrac}
\usepackage{placeins}
\usepackage[colorlinks]{hyperref}

\usepackage[english]{babel}
\graphicspath{{./}{Figures/}}

\usepackage{amssymb}
\usepackage{amsmath}
\usepackage{graphicx}
\usepackage[usenames,dvipsnames]{xcolor}
\usepackage{tikz}
\usepackage{wasysym}

\usepackage{algorithm2e}[ruled]

\newcommand{\vect}[1]{\vec{#1}}
\newcommand{\tens}[1]{\overline{\overline{#1}}}
\newcommand{\avg}[1]{\left\langle {#1} \right\rangle}

\newcommand{\Vc}[1]{Voronoi cell}
\newcommand{\Vcs}[1]{Voronoi cells}

\makeatletter
\def\ps@pprintTitle{%
  \let\@oddhead\@empty
  \let\@evenhead\@empty
  \def\@oddfoot{\reset@font\hfil\thepage\hfil}
  \let\@evenfoot\@oddfoot
}
\makeatother

\begin{document}

\begin{frontmatter}



\title{Microstructure-based prediction of hydrodynamic forces in stationary particle assemblies}

\affiliation[inst1]{organization={Lehrstuhl für Mechanische Verfahrenstechnik, Fakultät für Verfahrens- und Systemtechnik},
           addressline={Otto-von-Guericke-Universität}, 
            city={Magdeburg},
            postcode={39106}, 
            state={Sachsen-Anhalt},
            country={Germany}}

\author[inst1]{Berend van Wachem}
\author[inst1]{Hani Elmestikawy}
\author[inst1]{Victor Chéron}
\begin{abstract}
In the work, we derive novel hydrodynamic force models to describe the interaction of a flow with particles in an assembly when only an averaged resolution of the flow is available. These force models are able to predict the average drag on the particle assembly, as well as the deviations from the average drag force and the lift force for each individual particle in the assembly. 
To achieve this, PR-DNS of various particle assemblies and flow regimes are carried out, varying the particle volume fraction up to 0.6, and the mean particle flow Reynolds number up to 300.
To characterize the structure of the particles in the assembly, a Voronoi tessellation is carried out, and a number of scalars, vectors and tensors are defined based upon this tessellation. The microstructure informed hydrodynamic force models are based on symbolic regressions of these quantities derived from the Voronoi tessellation, the global particle volume fraction of the particle assembly and the flow regime represented by the Reynolds number, and the forces on the individual particles in the assembly.

The resulting hydrodynamic force models are single expressions can be directly employed in a Lagrangian particle tracking (LPT) or computational fluid dynamics/discrete element model (CFD/DEM) framework.
By comparing the results of the newly proposed hydrodynamic force models with an averaged force model, as is usually adopted in Lagrangian particle tracking simulations, we show that a significant increase in accuracy can be achieved, without significantly increasing the cost of the simulation.
\end{abstract}





\end{frontmatter}


\section{Introduction}
\label{sec:Introduction}
Particle-laden flows are widely encountered in environmental phenomena, such as transport of river sediments or avalanches, as well as in industrial applications, such as pneumatic conveying and fluidized beds. Such flows exhibit complex dynamics, which are ascribed to the wide range of time and length scales present in such flows, and the complex phenomena occurring at these various scales. Despite the large body of research into particle-laden flows, the reliable and affordable prediction of such flows is still lacking.

Particle-laden flows are commonly simulated by considering the particles as either a continuous Eulerian phase, or as discrete Lagrangian elements~\cite{Balachandar2010,Curtis2004}. A Eulerian description of the particles is capable of simulating a number of particles comparable to that of a complete system, but at the expense of the lack of information on the individual behavior and trajectories of the particles. On the other hand, a Lagrangian model can accurately describe the motion of each individual particle, but comes with a significant computational cost.
There are also various frameworks in which the continuous phase in particle-laden flows can be predicted, \textit{i.e.} direct numerical simulation (DNS), large eddy simulation (LES), and the Reynolds averaged Navier-Stokes (RANS) framework.

Another important aspect of the simulation framework applied to predict particle-laden flows, is the resolution of the particle in terms of the discretization spacing of the equations governing the fluid, typically referred to as the particle size to fluid mesh spacing ratio. There is currently a large body of research on \textit{fully resolved} particle-laden flow simulations~\cite[\textit{e.g.}][]{Uhlmann2005}. In such frameworks, the particles are resolved in the Lagrangian framework and the fluid is resolved with DNS. This implies that the boundary condition between the particle surface and the fluid flow is fully resolved. This method is also commonly referred to as particle resolved (PR) DNS, or PR-DNS.
The advantage of this approach, is that there are hardly any empirical parameters, as all important length- and time-scales are resolved by the methodology. However, this methodology comes with an excessive computational cost, even for a very limited number of particles in the flow, and is only feasible for low to moderate Reynolds number flows. This renders this framework unsuitable for predicting the behavior of most realistic particle-laden flows, although it can be used to study very small, typically academic, cases to obtain specific information from.

In most particle-laden flow simulation frameworks, the particle surfaces are not fully resolved, and the interaction between the fluid and the surfaces of the individual particles must be modelled~\cite{Anderson1967}. 
This hold for both the Eulerian description of the particle phase~\cite{Simonin1993}, as well as the Lagrangian description of the particle phase~\cite{Tsuji1987}.
Fluid-particle interaction models have been the focus of attention in the literature, by considering the pressure drop over a large number of particles~\cite{Wen1966,Ergun1952,Rowe1961}, by considering a theoretical analysis of a Stokes flow around a single particle~\cite{Maxey1983,Gatignol1983}, or by performing PR-DNS of one or a few particles in a numerical framework~\cite{Hill2001,Tenneti2011,Tenneti2014}.

When researching fluid-particle interaction models,
most of the time only the drag force is considered.
Some of these drag force correlations are summarized in Table \ref{tab:IsotropicDragModels}, which shows the multiplication factor of the actual drag force compared to the drag on a single particle in an infinite Stokes flow. 
These drag models predict the drag force based on two parameters of the particle-laden flow system: the mean flow Reynolds number, $\text{Re}_\textup{m}$, and the particle volume fraction of the particle assembly, $\varepsilon_{\mathrm{p}}$. These drag models were developed by numerous physical experiments~\cite{Ergun1952,Wen1966} as well as simulations~\cite{Hill2001b,Beetstra2007,Beetstra2007a,Tenneti2011,Tang2015}, and they are applicable to wide range of different particle volume fractions and flow regimes, characterized by the particle Reynolds number.
Because these models depend only on the volume fraction of the particle assembly, which needs to be determined over a large volume compared to the volume of an individual particle, as well as the \textit{mean} particle Reynolds number, which is also typically determined at a scale larger than an individual particle, these drag models can be considered to be isotropic, as no information on the local microstructure of the particle assembly is taken into account.
While there is no denying the ability of these drag models to predict the mean value of the drag force exerted on the particles in an assembly, isotropic drag closure models remain limited by the fundamental assumptions on which they rely, namely that the variation in spatial distribution of the particles is solely accounted for by the scalar entity, the volume fraction, and, consequently, that all particles in an assembly in a region with a similar relative velocity experience the same, or a very similar, drag force. 
\AtBeginEnvironment{tabular}{\small}
\begin{table}[h]
    \centering
    \begin{tabular}{|p{3cm}|p{7cm}|p{4cm}|}
    \hline
        \textbf{Reference} & \textbf{Drag Expression relative to Stokes drag} & \textbf{Description} \\
    \hline
         Ergun \cite{Ergun1952}& \begin{equation*}
             F(\varepsilon_{\mathrm{p}},\mathrm{Re_m}) = \frac{150 \varepsilon_{\mathrm{p}}}{10(1-\varepsilon_{\mathrm{p}})^2} + \frac{1.75 \mathrm{Re_m}}{1-\varepsilon_{\mathrm{p}}}
         \end{equation*}& Developed from experiments in fluidized experiments, applicable only for dense packing and $\mathrm{Re_m} < 1000$ \\
    \hline
         Hill et al. \cite{Hill2001b} & \vspace{-4mm}{\begin{align*}
             F(\varepsilon_{\mathrm{p}},\mathrm{Re_m}) & = \frac{150 \varepsilon_{\mathrm{p}}}{10(1-\varepsilon_{\mathrm{p}})^2} + \frac{1}
             {6\varepsilon_{\mathrm{p}} - 10 \varepsilon_{\mathrm{p}}^2}{(1- \varepsilon_{\mathrm{p}})^2}\\ 
         \end{align*}} & Developed from Lattice Boltzmann simulations, applicable for $0.1\leq \varepsilon_{\mathrm{p}} \leq 0.6$ and $\mathrm{Re_m}\leq 200$, mono-disperse assemblies \\
    \hline
         Beetstra et al.\cite{Beetstra2007a} & \vspace{-4mm}{\begin{align*}
              F(\varepsilon_{\mathrm{p}},\mathrm{Re_m}) &= \frac{10 \varepsilon_{\mathrm{p}}}{(1-\varepsilon_{\mathrm{p}}^2)} + (1-\varepsilon_{\mathrm{p}})^2(1+1.15\varepsilon_{\mathrm{p}}^{\frac{1}{2}})\\ & + \frac{0.413\mathrm{Re_m}}{24(1-\varepsilon_{\mathrm{p}})^2} \left[ \frac{(1-\varepsilon_{\mathrm{p}})^{-1} 3\varepsilon_{\mathrm{p}}(1-\varepsilon_{\mathrm{p}}) +8.4 \mathrm{Re_m}^{-0.343}}{1+10^{3\varepsilon_{\mathrm{p}}} \mathrm{Re_m}^{-(1+4\varepsilon_{\mathrm{p}})/2}}\right]
         \end{align*}}& Developed from Lattice Boltzmann simulations, applicable for $0.1\leq \varepsilon_{\mathrm{p}} \leq 0.65$ and $\mathrm{Re_m}\leq 1000$, mono-disperse and bi-disperse assemblies \\
    \hline
        Tenneti et al. \cite{Tenneti2011} & \vspace{-4mm}{\begin{align*}
            F(\varepsilon_{\mathrm{p}}, \mathrm{Re_m}) &= \frac{1 +0.15 \mathrm{Re_m}^{(0.687)}}{(1-\varepsilon_{\mathrm{p}})^3} + \frac{5.81\varepsilon_{\mathrm{p}}}{(1-\varepsilon_{\mathrm{p}})^3}+ \frac{0.48 \varepsilon_{\mathrm{p}}^{\frac{1}{3}}}{(1-\varepsilon_{\mathrm{p}})^4}\\ &+\varepsilon_{\mathrm{p}}^3 \mathrm{Re_m}\left[ 0.95 + \frac{0.61\varepsilon_{\mathrm{p}}^3}{(1-\varepsilon_{\mathrm{p}})^2}\right]
        \end{align*}} & Developed from immersed boundary method simulations; applicable for $0.1\leq \varepsilon_{\mathrm{p}}\leq 0.5$ and $ \mathrm{Re_m}\leq 300$, monodispersed assemblies\\
    \hline
        Tang et al. \cite{Tang2015}& \vspace{-4mm} {\begin{align*}
            F(\varepsilon_{\mathrm{p}}, \mathrm{Re_m}) & = \frac{10 \varepsilon_{\mathrm{p}}}{(1-\varepsilon_{\mathrm{p}})^2} + (1-\varepsilon_{\mathrm{p}})^2 (1+1.5\varepsilon_{\mathrm{p}}^{\frac{1}{2}}) \\
            &+ \mathrm{Re_m}\left[ 0.11 \varepsilon_{\mathrm{p}} (1+\varepsilon_{\mathrm{p}}) - \frac{0.00456}{(1-\varepsilon_{\mathrm{p}})^4} \right. \\ &\left. + \mathrm{Re_m}^{-0.343}\left( 0.169 \, (1-\varepsilon_{\mathrm{p}}) + \frac{0.0644}{(1-\varepsilon_{\mathrm{p}})^4}\right)\right]
        \end{align*}} & Developed from immersed boundary method simulations; applicable for $0.1\leq \varepsilon_{\mathrm{p}}\leq 0.6$ and $ \mathrm{Re_m}\leq 1000$, monodispersed assemblies  \\
    \hline
    \end{tabular}
    \caption{Summary of some isotropic drag models from the literature. The drag expression is expressed as the multiplication factor of the drag force on a particle compared to the Stokes drag as a function of the particle volume fraction and the mean flow Reynolds number.}
    \label{tab:IsotropicDragModels}
\end{table}

In~\citet{Knight2020}, PR-DNS simulations of the flow between polydisperse particles in dense assemblies were carried out. It was shown, that existing drag isotropic drag models poorly capture the forces on the individual particles in the assembly, and that the forces on the individual particles has a much wider spread, than the forces predicted by the isotropic drag models for these particles. In their work, it was suggested to carry out a Voronoi tessellation of the particle assembly, in order to obtain information on the microstructure of the particle assembly that can be used to enhance the prediction of the forces on the individual particles.
This was carried out in \citet{Che2021}, in which a radical Voronoi tessellation was carried out of a particle assembly, and the volume fractions of the individual Voronoi cells, the so-called \textit{local} volume fractions, were used to predict the forces on the individual particles, using existing isotropic drag correlations. Because this method is grid independent and there are no empirical parameters required to construct the Voronoi grid, the resulting smooth local volume fraction field is unique, and results in a more accurate estimate of the drag force on the individual particles, even if the local expression to determine the drag force is still an isotropic one.

In reality, however, the forces experienced by the individual particles within random assemblies deviates substantially from the average value predicted by an isotropic drag correlation. This deviation is mainly attributed to the random spatial distribution of the particles. To explore this, \citet{Akiki2016} used  PR-DNS to reveal that the upstream, downstream and lateral positions of neighbouring particles have distinct influences on the drag and lift forces on each individual particle, compared to the mean forces on the assembly as a whole. Therefore, they suggested introducing a composite parameter reflecting the relative positions of neighbouring particles to characterize the anisotropy. 
After this, a number of other works studied the effect of the arrangement in the particle assembly on the distribution of the forces on the individual particles in the assembly, ranging from deterministic~\cite{Akiki2017}, to stochastic~\cite{Seyed-Ahmadi2020}, to machine-learning~\cite{Siddani2023} based models. All these models make use of the particle positions and spatial distribution, information readily available in Lagrangian particle simulations, but differ in their assumptions, methodology and predictive capabilities.

The pairwise interaction extended point-particle (PIEP) model~\cite{Akiki2017} quantifies the modification of the drag force on a particle due to the presence of other particles
in a two-step process: firstly, a pairwise particle interaction assumption is made, from which the undisturbed flow around a reference particle, \textit{i.e.} the flow solution in case of the absence of that particle, caused by the influence each neighbour, is estimated. Each contribution of individual neighbouring particle to this undisturbed flow is considered separately, and the overall flow field obtain by linear superposition of each neighbouring particle. The force associated with this undisturbed flow is then estimated by using a generalized form of Faxén theorem~\cite{Gatignol1983,Maxey1983}. An extension of their model to determine the torque as well was derived in \cite{Akiki2017a}. This necessitates a definition of the so-called particle ``neighbourhood'', which guarantees an accurate estimation of the drag forces. The results show a significant reduction in drag error when 15-40 particles are considered. This error reduction, however, decreases with increasing Reynolds number and volume fraction, with the effect of the volume fraction being more prominent.

The microstructure-informed probability driven point-particle (MPP) model~\cite{Seyed-Ahmadi2020} is a data-driven model, which is based on statistical information gained from PR-DNS simulations of particle-laden flows. It relies on the idea that particles with similar neighbourhoods are likely to have similar drag and lift force distributions. Consequently, the probability density functions (PDF) of the neighbourhood spatial distributions and those of the force variation exhibit some correlation and can be employed to derive a model. 
This model distinguishes between positive and negative relative forces, and applies a least-square linear regression to the system spanned by a PDF of the spatial distributions and that of the actual force variations obtained from PR-DNS results to derive model coefficients, while assigning different probabilities to particle relative positions. 
The neighbourhood which the model takes into account includes the 30 closest particles, and its performance shows remarkable predictive capabilities, although the accuracy reduces as the particle Reynolds number and/or the particle volume fraction is increased.

In the recent work of \citet{Hardy2022}, the microstructure of the solid phase is characterized by a weighted linear superposition of the tensorial quantities which describe the pairwise particle interactions. These quantities are based on the so-called fabric tensor used to calculate the stresses in porous media~\cite{Oda1982}. A Gaussian filter is used to assign different weights to neighbouring particles, which decreases with increasing inter-particle distance. In addition, an asymmetrical vector entity is considered, which takes into account the relative orientation and relative flow velocity direction~\cite{Oda1982}.  The total number of neighbouring particles included in that structure is 216. A sensitivity analysis carried out in their work, however, shows that 64 neighbours yield satisfactory results. The statistics of these quantities were studied in detail and their mutual dependence is verified. The verified existing correlations to drag force variations allowed the derivation of a multi-linear model capable of predicting variation in the forces on the particles in both streamwise and transverse flow directions. Similar to the PIEP and MPP models, the model performs best at low Reynolds numbers and low volume fractions. As these quantities are increased, the predictive capability of the model deteriorates, and the values of the forces are underestimated. To correct for this, a Reynolds number dependent stochastic term is added. Moreover, although there are single forms of the expressions to predict the drag and lift forces on the particles, the empirical coefficients in these expressions are significantly different for the three mean Reynolds numbers and four volume fractions studied in their work, making the application in a Lagrangian particle tracking (LPT) or CFD/DEM simulation impractical. 

A different methodology is presented in the contribution of \citet{Seyed-Ahmadi2021}, in which they use machine learning techniques to develop a data-driven drag model. This physics-inspired neural-network model (PINN) offers a great versatility in the mapping of complex nonlinear interactions, without making strong prior assumptions about the model structure. Under the assumption of pairwise interaction, the model is able to reach results of the same order of accuracy of that shown in PIEP, MPP and fabric-tensor-based model, while including no more than 10 particles in the arbitrarily defined neighbourhood domain. This means, that characterizing the structure of a relatively small neighbourhood of particles around a particle suffices to predict the variations in drag force expected on this particle.

In this work, we derive novel hydrodynamic force models to describe the interaction of a flow with particles in an assembly. These force models are able to predict the average drag on the particle assembly, as well as the deviations from the average drag force and the lift force for each particle in the assembly. The hydrodynamic force models are single expressions, are based on the variables describing the microstructure of the particle assembly and the fluid flow regime, and can be directly employed in a LPT or CFD/DEM framework.
To achieve this, PR-DNS of various particle assemblies and flow regimes are carried out, varying the particle volume fraction between 0.1 and 0.6, and the mean particle flow Reynolds number between 0.1 and 300.
To characterize the structure of the particles in the assembly, a Voronoi tessellation is carried out, and a number of scalars, vectors and tensors are defined based upon this tessellation. The microstructure informed drag model is based on the correlation of these quantities derived from the Voronoi tessellation, the global particle volume fraction of the assembly and the flow regime represented by the Reynolds number, and the forces on the individual particles in the assembly.

This paper is structured as follows. In section 2, we briefly summarize the numerical framework for the PR-DNS simulations of the particle assemblies. In section 3, we outline the Voronoi tessellation and the scalar, vector and tensor quantities which are based on this tessellation. In section 4, we present the simulation setup and validation, and in section 5 we present the results and the resulting microstructure informed drag force model regression. Finally, in section 6 conclusions are drawn.
\section{PR-DNS framework}
\label{sec:HyBM method}
For the particle-resolved direct numerical simulations (PR-DNS), we apply the smooth immersed boundary method (IBM)~\citep{Uhlmann2005}, using the direct-forcing formulation of~\citet{AbdolAzis2019} with the additional correction of~\citet{Cheron2023a}, to enable the accurate flow modelling between two near particle surfaces.
The IBM couples the discretized equations governing the fluid flow in the domain, described in an Eulerian framework, and the representation of the surfaces of the particles with a Lagrangian framework. The Lagrangian framework consists of evenly spaced markers, called Lagrangian markers, discretizing the surfaces of each of the particles.

The incompressible fluid phase is governed by the Navier-Stokes equations with an additional source term, which are written as
\begin{align}
    \rho_\mathrm{f} \left(\frac{\partial   \vect{u}}{\partial{t}} + \left(\vect{u} \cdot \nabla \right) \vect{u} \right) & = - \nabla p + \mu \nabla^2\vect{u}+\vect{f} \\
\nabla \cdot \vect{u} & = 0
\end{align}
where $\rho_\mathrm{f}$ is the density of the fluid, $\vect{u}$ is the fluid velocity vector, $p$ is the pressure, $\mu$ the dynamic viscosity of the fluid, and $\vect{f}$ represents the momentum source arising from the presence of the immersed boundaries and other body forces.
Applying a finite-volume discretization in a collocated grid arrangement, the discretized momentum equation for the $i$-th Eulerian cell, spanning the volume $\Omega_i$, can be summarized as
\begin{equation}\label{eq:NSeq_Mom_Discret}
 \int_{\Omega_i} \rho_\mathrm{f} \dfrac{\partial  \vect{u}_i}{\partial t} \ \mathrm{d} V 
+ \vect{c}_i = -\vect{b}_i + \vect{d}_i + \vect{f}_i
\end{equation}
where the first term of the left-hand-side is the transient term, and $\vect{c}_i $, $\vect{b}_i$, $\vect{d}_i$, $\vect{f}_i$ are the discretized advection term, pressure term, diffusion term and the source term arising from the presence of the immersed boundary and body forces, respectively. 
The Eulerian framework is based on a fully-coupled finite-volume framework with a collocated variable arrangement~\cite{Bartholomew2018,Denner2020}.

As the Eulerian and Lagrangian frameworks are independent, 
the necessity to modify or reconstruct a boundary-conformal Eulerian grid to match the Eulerian one is circumvented, at the expense that an explicit interpolation/spreading strategy is required to connect the Eulerian and Lagrangian fluid variables~\citep{Peskin2003}.
The required interpolation and spreading operators are constructed through a mollified moving-least-squares algorithm~\citep{Bale2021}.
The size of the operator, in terms of Eulerian mesh cell widths, determines the number of fluid cells used for the interpolation and the spreading of the fluid variables.
In this work, a five-point spline kernel function is used~\citep{Bao2016}, and the spreading of the source terms arising from the coupling of the particle surfaces to the fluid momentum equations is scaled by a relaxation factor~\citep{Zhou2021}.
This relaxation factor is based on stability condition criterion, and controls the rate at which the no-slip condition is reached as well as the magnitude of the no-slip error.
The direct-forcing approach~\cite{AbdolAzis2019,Cheron2023a} is used to compute the feedback force for each $j$-th Lagrangian marker, which reads as
\begin{equation}
    \vect{F}_{j}^n = \frac{\rho_\mathrm{f}}{\Delta t} \left( \vect{U}_{\text{IB},j}^n - \vect{U}_j^{n-1}\right) + \vect{C}_j^n + \vect{B}_j^n - \vect{D}_j^n 
    \label{eq:LagForce}
\end{equation}
where the super-script $n$ denotes the time level at which the quantities are to be evaluated. 
The symbols in capital indicate the variables which are interpolated from the Eulerian framework to the Lagrangian markers, where $\vect{U}_{\text{IB},j}$ is the prescribed velocity vector of the $j$-th Lagrangian marker, and $\vect{U}_j$,  $\vect{C}_j$,  $\vect{B}_j$,  and $\vect{D}_j$, are the interpolated velocity, advection, pressure, and diffusion terms of the governing momentum equations, respectively. 

\begin{figure}[htbp!]
\centering
\includegraphics[width=0.485\textwidth]{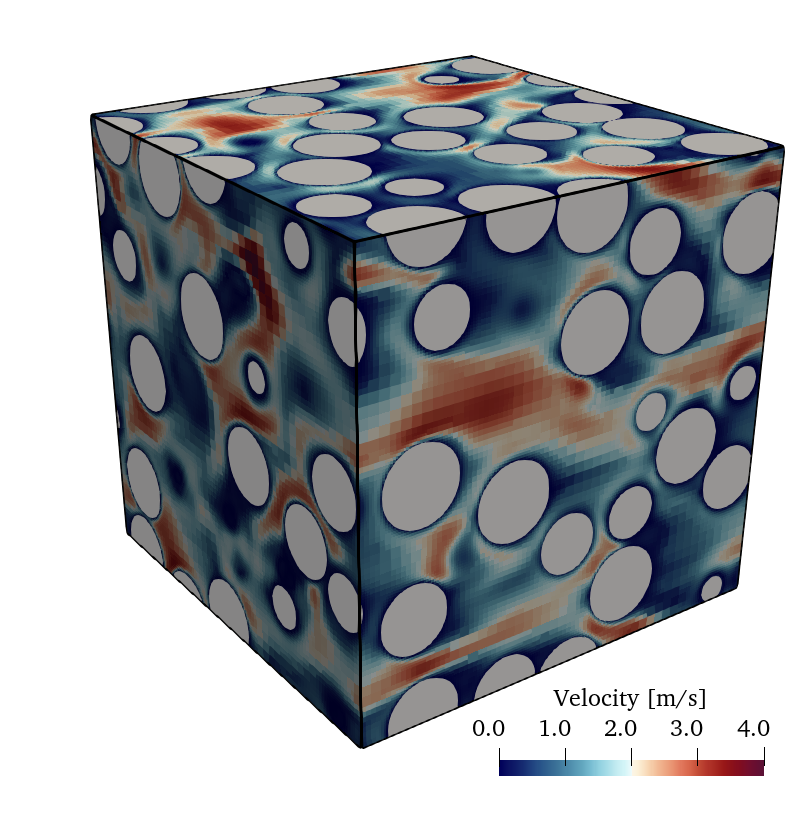}
\includegraphics[width=0.485\textwidth]{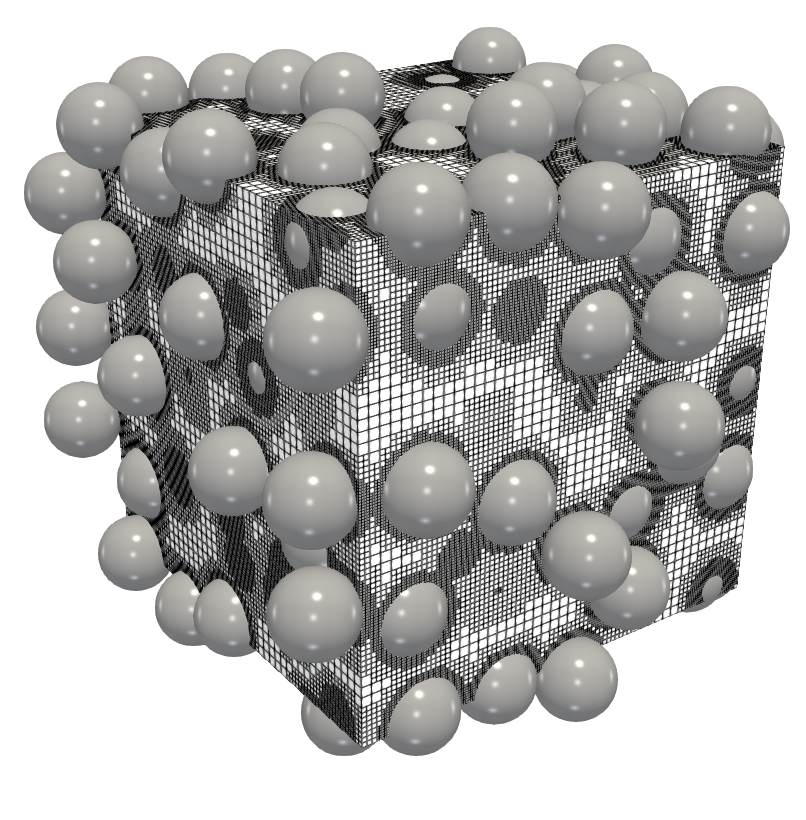}
\caption{A snapshot of the fluid flow through a periodic box of a random assembly of particles (left) with the adapted fluid mesh (right).}
\label{fig:PRDNS1}
\end{figure}

To reduce the computational cost, the DNS are performed with a dynamic adaptive mesh refinement (AMR) for the Eulerian fluid mesh, in order to focus the computational efforts in the regions with large gradients in the fluid velocity, \textit{e.g.} near the particle surfaces and near vortices in the flow.
The refinement and coarsening criteria are a distance-based refinement criterion and a vorticity-based criterion, which ensure that the fluid mesh is refined near the surface of the particle and in the wake of the particle. An instantaneous snapshot of the fluid flow through a periodic box with a random assembly of particles is shown in Figure~\ref{fig:PRDNS1}. 
 \subsection{Determining the hydrodynamic forces from the PR-DNS}
The PR-DNS are carried out in a periodic cube, with volume $V$, with $N_\mathrm{p}$ non-overlapping monodispersed spherical particles. 
The periodic cube is divided into $N_\mathrm{f}$ Eulerian fluid cells, in which the discretized equations governing the fluid flow are solved to determine the fluid flow properties.
Each Eulerian fluid cell in the cube can be occupied by fluid, by the particle, or partly by the fluid and partly by the particle. The flow in the periodic cube is driven by specifying an external pressure drop, $\delta P$, by adding this source term to the fluid momentum equations as an additional body force in the $x$-direction.

To determine the amount of fluid in each Eulerian mesh cell, an indicator function is introduced, $\mathcal{I}_\mathrm{f}(\vect{x})$, as:
\begin{equation}
    \mathcal{I}_\mathrm{f}(\vect{x}) = 
        \begin{cases}
        1, & \text{if $\vect{x}$ is in the fluid phase,}\\
        0, & \text{otherwise}
        \end{cases}
\end{equation}
The fluid volume fraction from the Eulerian cell $i$, $\varepsilon_{\mathrm{f},i}$, can be determined as
\begin{equation}
    \varepsilon_{\mathrm{f},i} = \frac{1}{\Omega_i} \int\limits_{\Omega_i} \mathcal{I}_\mathrm{f}(\vect{x}) d\vect{x}
\end{equation}
where $\Omega_i$ is the volume of Eulerian cell $i$. 
The volume integral is evaluated by means of a second-order accurate mid-point rule~\cite{Kempe2012}.
The averaged fluid phase velocity in the cube is computed as the spatial average over the volume of the cube~\cite{Zick1982}: 
\begin{equation}
    \avg{U} = \frac{\int\limits_{V} \mathcal{I}_\mathrm{f}(\vect{x}) \vect{u}(\vect{x}) d\vect{x}}{\int\limits_V \mathcal{I}_\mathrm{f}(\vect{x}) d \vect{x}}
\end{equation}
where $V$ is the volume of the cube.
The overall particle volume fraction in the cube can be determined by dividing the total volume of the particles by the volume of the cube,
\begin{equation}
    {\varepsilon_{\mathrm{p}}} = \dfrac{ N_\mathrm{p} \pi d_\mathrm{p}^3}{6 V} 
\end{equation}
The overall fluid volume fraction can be determined by averaging over $\varepsilon_{\mathrm{f},i}$, or by using the definition
${\varepsilon_{\mathrm{f}}} = 1 - {\varepsilon_{\mathrm{p}}}$.
The mean flow Reynolds number is computed with the average fluid phase velocity as
\begin{equation}
    \mathrm{Re_m} = \frac{ {\varepsilon_{\mathrm{f}}} \avg{U} d_\mathrm{p}}{\nu_\mathrm{f}}
\end{equation}
where $d_\mathrm{p}$ is the diameter of the particle, and $\nu_\mathrm{f}$ is the kinematic viscosity of the fluid.

From the PR-DNS simulations, the force on particle $k$ exerted by the fluid is given by
\begin{equation}
\vect{F}_{\mathrm{p},k} = - \rho_\mathrm{f} \sum\limits_{\forall (j \in  k)} \vect{F}_{j,k} \mathcal{V}_{j,k} + \rho_\mathrm{f} \dfrac{d}{dt}\int\limits_{V_{\mathrm{p},k}} \vect{u}(\vect{x}) d\vect{x}\ - {\delta P} \hat{x} \, V_{\mathrm{p},k}
\label{eq:PR-DNS-Force}
\end{equation}
where $V_{\mathrm{p},k}$ is the volume of particle $k$ and $\sum\limits_{\forall (j \in  k)}$ indicates a summation over all Lagrangian markers $j$ of particle $k$.
In the above equation the first term represents the IBM feedback forces, with $\mathcal{V}_{j,k} $ the Lagrangian weight associated to the $j$-th marker of the $k$-th particle, and $\vect{F}_{k,j}$ is the force associated with the Lagrangian marker $j$ of particle $k$ and is determined from Eq.~\eqref{eq:LagForce}.
The second term represents the rate of change in fluid momentum inside the particle, which is fictitious and should be subtracted.
The last term is the mean pressure gradient driving the flow~\cite{Zhao2013a} in the $x$-direction, which needs to be accounted for in the force computation.

\section{Microstructure characterization}

\subsection{Voronoi tessellation}
\label{subsec: Voronoi}
\begin{figure}[htbp!]
\centering
\includegraphics[width=0.8\textwidth]{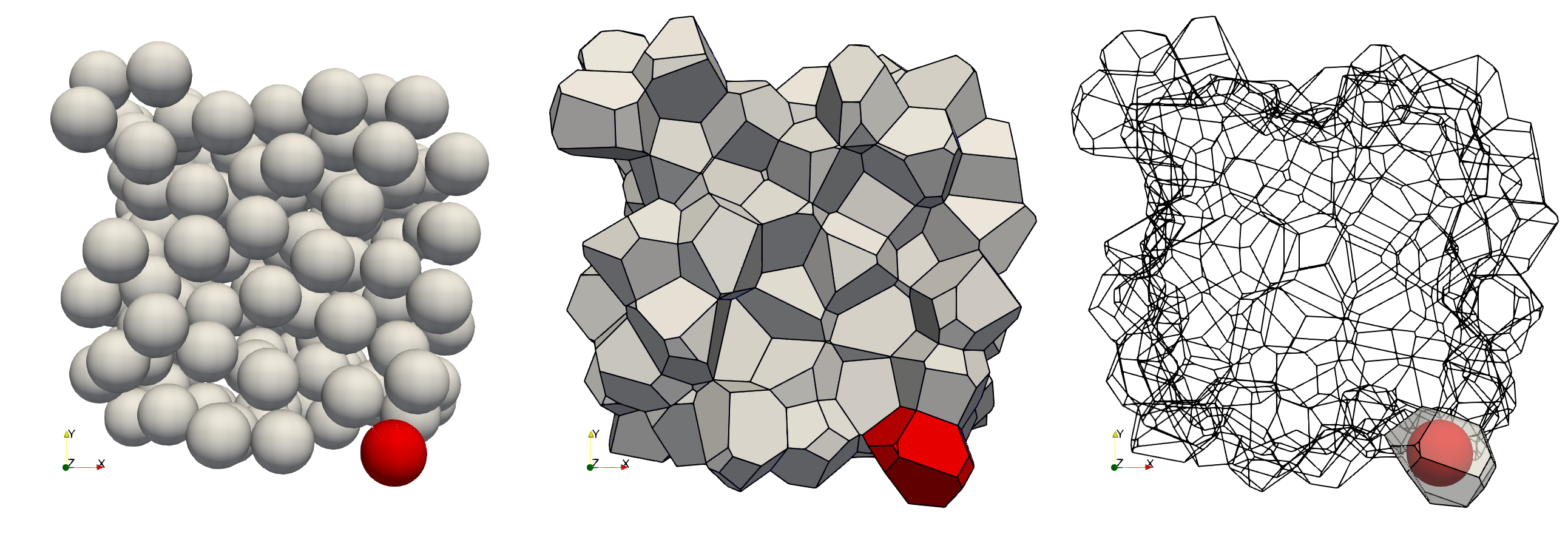}
\caption{A random monodispersed  assembly of particles with a solid volume fraction ${\varepsilon_{\mathrm{p}}} = 0.4$ (left), and the Voronoi tessellation (middle) and focusing on 1 particle in red (right).}
\label{fig:voronoi-diagram-1}
\end{figure}

Voronoi tessellation is a useful tool to characterize the structure of particle assemblies in granular flows, as it provides a unique microscopic description of the arrangement of the particles in the assembly. It can be used, for instance, to calculate the local solid volume fraction of a particle in a Voronoi volume to determine the porosity of a material~\citep{Che2021}, or quantify the anisotropy in the distribution of particles~\citep{Monchaux2010}.\\
In the context of monodispersed assembly of spheres, the Voronoi tessellation consists of partitioning the space into cells around each particle. These Voronoi cells are polyhedra, which represent the space surrounding each particle. Each Voronoi cell is constructed by connecting the centers of neighbouring particles, and placing the face of the polyhedron perpendicular to this connection at the midpoint between the nearest neighbouring particles.
The voro++ library~\cite{Rycroft2009}, based on a Delaunay triangulation method, is used to construct the Voronoi tessellation in our framework.
An example of the Voronoi tessellation of monodispersed assemblies of fixed spherical particles of solid volume fraction ${\varepsilon_{\mathrm{p}}} = 0.4$ is shown in figure~\ref{fig:voronoi-diagram-1}.

\subsection{Minkowski tensors}
\label{subsec: Minkowski}
From the Voronoi tessellation of a particle assembly, the Minkowski tensors, also more generally referred to as Minkowski functionals, are used to capture the anisotropic and non-local properties of the particle assembly.
These tensors are directly calculated from the Voronoi tessellation, and represent a set of quantities which describe the shape and size of a geometrical object in terms of its volume, surface area, and curvature.
As discussed in~\citet{SchroederTurk2013}, several equivalencies exist between the Voronoi tessellation and the Minkowski tensors, for instance the zeroth order Minkowski tensors represent the volume and surface area of the Voronoi cell.

The Minkowski tensors which are used in this work to characterize the structure of the local assembly are given in table~\ref{tab:MinkowskiTensors}, and are discussed below in a proposed dimensionless form.
To simplify the notation, the subscript $k$, which indicates the $k$-th particle in the assembly, is dropped and all tensors quantities are defined for each individual particle, or Voronoi cell, of the array.

\AtBeginEnvironment{tabular}{\small}
\begin{table}[h]
    \centering
    \begin{tabular}{lcc}
        \multicolumn{2}{c}{\em{~Tensors rank 0}} & unit\\
        \hline\\[-2mm]
        $W_{\textup{0}}$ & $\int_\mathrm{K} d V_\mathrm{K}$ & $[m^3]$\\[2mm]
        $W_{\textup{1}}$ & $\int_{A_\mathrm{K}} d A_\mathrm{K}$ & $[m^2]$\\[3mm]
        \multicolumn{2}{c}{\em{~Tensors rank 1}} & unit\\
        \hline\\[-2mm]
        $\vect{W}^{\textup{1,0}}_{\textup{0}}$ & $\int_\mathrm{K} \vect{x} d V_\mathrm{K}$ & $[m^4]$\\[2mm]
        $\vect{W}^{\textup{1,0}}_{\textup{1}}$ & $\frac{1}{3}\int_{A_\mathrm{K}} \vect{x} d A_\mathrm{K}$ & $[m^3]$\\[3mm]
        \multicolumn{2}{c}{\em{~Tensors rank 2}} & unit\\
        \hline\\[-2mm]
        $\tens{W}^{\textup{2,0}}_{\textup{0}}$ & $\int_\mathrm{K} (\vect{x} \otimes \vect{x}) d V_\mathrm{K}$ & $[m^5]$\\[2mm]
        $\tens{W}^{\textup{2,0}}_{\textup{1}}$  & $\frac{1}{3}\int_{A_\mathrm{K}} (\vect{x} \otimes \vect{x}) d A_\mathrm{K}$ & $[m^4]$\\[2mm]
        $\tens{W}^{\textup{0,2}}_{\textup{1}}$  & $\frac{1}{3}\int_{A_\mathrm{K}} (\vect{n} \otimes \vect{n}) d A_\mathrm{K}$ & $[m^2]$\\[2mm]
    \end{tabular}
    \caption{Definition of the Minkowski tensors of rank 0 (scalars), rank 1 (vectors), and rank 2 (tensors). These tensors are obtained from volume integral over the volume of the Voronoi cell, $V_\mathrm{K}$, and the surface integral over the surface of the Voronoi cell, $A_\mathrm{K}$. $\vect{x}$ and $\vect{n}$ are both position and normal vectors given in the frame of reference centered at the center of the particle.
    \label{tab:MinkowskiTensors}}
\end{table}

The Minkowski tensors of rank 0 are referred to as the zeroth-order moments of the Minkowski functionals. 
They are defined based on the volume and the total surface area of the Voronoi cell, which are directly obtained from the Voronoi tessellation.
For example, $W_{\textup{0}}$ corresponds to the volume of the Voronoi cell, $V_\mathrm{K}$, and $W_{\textup{1}}$ to its total surface area.
From the knowledge of the volume of a \Vc{} and the diameter of the particles, the local solid volume fraction, $\varepsilon_{\mathrm{V}}$, of the particle associated with its unique \Vc{} can be defined as
\begin{equation}
    \varepsilon_{V} = \frac{ \frac{4}{3} \pi \big(\nicefrac{d_p}{2}\big)^3}{V_\mathrm{K}}
    \label{eq: Local Volume Fraction}
\end{equation}
The local solid volume fraction, $\varepsilon_{V}$, provides a measure of local density of the particles near the specific particle, which may deviate from the mean solid volume fraction of the whole particle assembly, if the particle assembly is inhomogeneous.
From the zeroth order tensor $W_{\textup{1}}$ we define a non-dimensional scalar quantity as
\begin{equation}
\varepsilon_{S} = \dfrac{W_{\textup{1}}}{\mathcal{S}_\mathrm{K}}
\end{equation}
where $W_{\textup{1}}$, obtained from the summation over the area of the faces bounding the \Vc{}, is scaled by a reference surface area $\mathcal{S}_\mathrm{K}$, defined as the surface area of a sphere with volume $V_\mathrm{K}$.

The Minkowski tensors of rank 1 provide vectorial quantities on the direction of the anisotropy of the Voronoi cell.
The tensor $\vect{W}^{\textup{1,0}}_{\textup{0}}$ and $\vect{W}^{\textup{1,0}}_{\textup{1}}$, given in table~\ref{tab:MinkowskiTensors}, are referred to as the center of mass and surface centroid vectors.
The center of mass is obtained for a Voronoi cell, assuming it is homogeneously filled with material of constant density, by multiplying the co-ordinates of the center of mass of the Voronoi cell in its reference frame by its volume. 
A dimensionless vector based on the volume of the Voronoi cell is defined as
\begin{equation}
\vect{\varepsilon}_{W^{\textup{1,0}}_{\textup{0}}} = \dfrac{\vect{W}^{\textup{1,0}}_{\textup{0}}}{V_\mathrm{K}^{\nicefrac{4}{3}}}
\end{equation}
Similarly, we define the surface centroid vector in a dimensionless form based on the volume of the Voronoi cell
\begin{equation}
\vect{\varepsilon}_{W^{\textup{1,0}}_{\textup{1}}} = \dfrac{\vect{W}^{\textup{1,0}}_{\textup{1}}}{V_\mathrm{K}}
\end{equation}

The Minkowski tensors of rank 2 provide detailed information on the degree of anisotropy of the Voronoi cell.
The tensor  $\tens{W}^{\textup{2,0}}_{\textup{0}}$, referred to as the solid moment tensor or moment of inertia, provides information on the orientation, elongation, and degree of anisotropy of the \Vc{}, and is defined in dimensionless form by
\begin{equation}
\tens{\varepsilon}_{W^{\textup{2,0}}_{\textup{0}}} = \frac{\tens{W}^{\textup{2,0}}_{\textup{0}}}{V_\mathrm{K}^{\nicefrac{5}{3}}}
    \label{eq: Solid Moment Tensor}
\end{equation}
where $\tens{W}^{\textup{2,0}}_{\textup{0}}$ is integrated over the volume of the Voronoi cell excluding the volume of the particle.
Two additional second order tensors are defined based on the surface of the Voronoi cell, the tensor  $\tens{W}^{\textup{2,0}}_{\textup{1}}$, also referred to as the hollow moment tensor, and defined in dimensionless form by
\begin{equation}
\tens{\varepsilon}_{W^{\textup{2,0}}_{\textup{1}}} = \frac{\tens{W}^{\textup{2,0}}_{\textup{1}}}{\left(W_{\textup{1}}\right)^{2}}
    \label{eq: Hollow Moment Tensor}
\end{equation}
Finally, the tensor $\tens{W}^{\textup{0,2}}_{\textup{1}}$ is the integral tensorial characterization of the normal vector distribution, and provides insight on the intrinsic anisotropy of the \Vc{}. This tensor in dimensionless form is defined as 
\begin{equation}
\tens{\varepsilon}_{W^{\textup{0,2}}_{\textup{1}}}= \frac{\tens{W}^{\textup{0,2}}_{\textup{1}}}{W_{\textup{1}}}
    \label{eq: Normal Distribution Tensor}
\end{equation}

The ratio of the smallest to the largest eigenvalue of a second order Minkowski tensor is defined as the anisotropic scalar value. For an isotropic cell, the value of this ratio is unity.
For the solid second order moment tensor this anisotropic scalar value is defined as
\begin{equation}
\left({\beta}^{\textup{2,0}}_{\textup{0}}\right) = \left(\frac{\min\vert\lambda^{\textup{2,0}}_{\textup{0}}\vert}{\max\vert\lambda^{\textup{2,0}}_{\textup{0}}\vert}\right)
\end{equation}
The definition of the anisotropic ratio $\beta$ can be made similarly for the other second order tensors, given by
Eq.~\eqref{eq: Hollow Moment Tensor}-\eqref{eq: Normal Distribution Tensor}.

\begin{figure}[htbp!]
\centering
\includegraphics[width=0.6\textwidth]{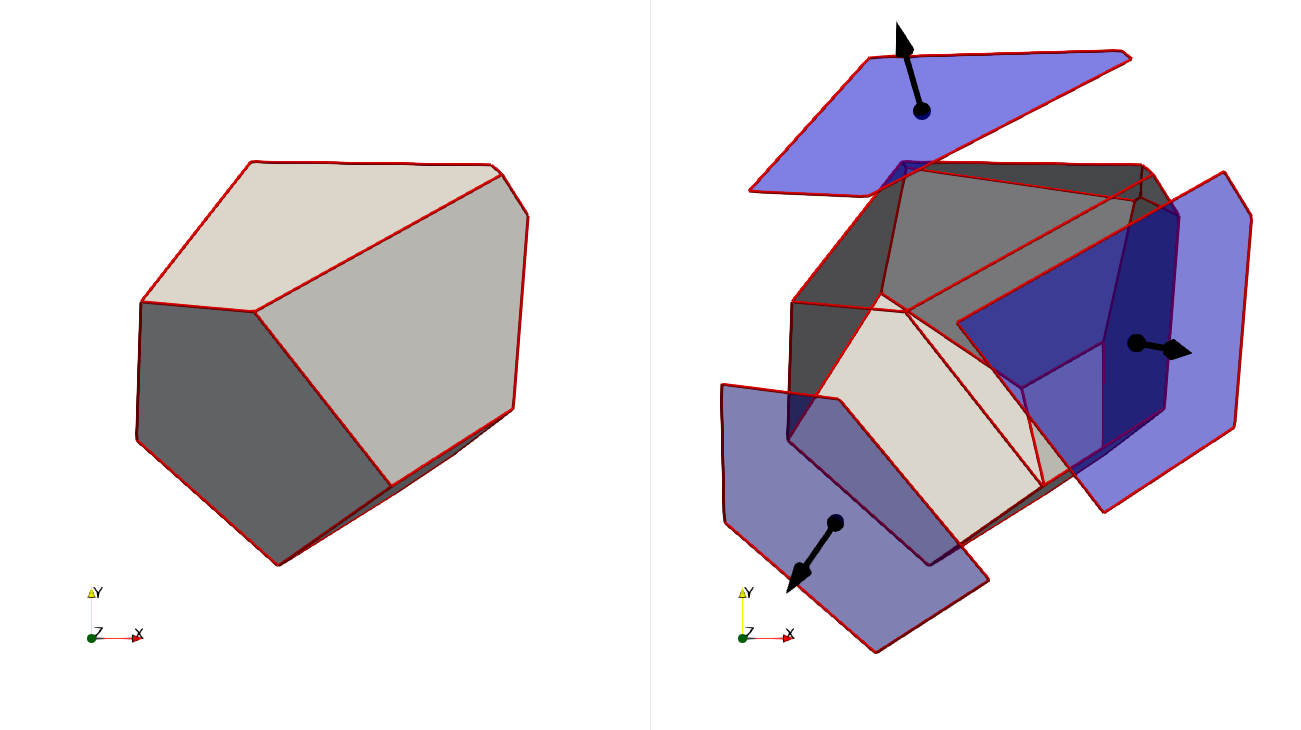}
\caption{Computation of the area, center, and normal of the faces embedding the \Vc{} from the list of vertices belonging to each face.}
\label{fig:voronoi-diagram-2}
\end{figure}

In addition to the Minkowski components, we define a vector representing the Voronoi cell-stretching, defined as the sum of the stretching vectors of each face, weighted by its area and scaled by the inverse of the distance from the center of the Voronoi cell to the face.
The cell stretching vector is defined in dimensionless form by
\begin{equation}
\vect{\mathcal{D}}  = \frac{\sum_{j}A_{j}\vect{n}_j{d_{j}^{-1}}}{\sum_{j} A_{j}d_{j}^{-1}}
    \label{eq: Vector cell stretching}
\end{equation}
where $\sum_{j}$ indicates a summation over all the faces of the Voronoi cell, $A_{j}$ is the area associated with the $j$-th face of the Voronoi cell, $\vect{n}_{kj}$ is the normal vector to the $j$-th face of the Voronoi cell, and $d_{j}$ is the distance from the center of the Voronoi cell $k$ to the $j$-th face of the Voronoi cell.
The computation of the normal $\vect{n}_{j}$ to the $j$-th face is obtained by the cross-product of the three successive vertices of face $j$, see figure~\ref{fig:voronoi-diagram-2}.
The center of the $j$-th face of the Voronoi cell is obtained from the geometric decomposition of the $n$-sided polygon into a total of $n-2$ triangles.

\section{PR-DNS simulation setup}
\label{sec: PR-DNS simulations}

The numerical framework presented in Section~\ref{sec:HyBM method} is employed to carry out the simulations of the fluid flow through random assemblies of fixed monodispersed spherical particles.
The PR-DNS are performed in a periodic cubic box of length $L$.
A Cartesian grid is used for discretizing the Eulerian fluid governing equations. This grid is dynamically refined or coarsened using adaptive mesh refinement (AMR).
The finest refinement level of the Eulerian grid is found at the interface between the fluid and the particles and in regions with a very high vorticity, as shown in figure~\ref{fig:PRDNS1}. 
The spatial resolution is determined by the finest level of refinement and is a function of the particle diameter. 
From our experience, a resolution of $d_p/\Delta x = 32$ leads to accurate results for the flow past particles~\cite{Cheron2023a}, an error in the drag force of less than 1\% is reached for simulations of fluid flow through random assemblies of large solid volume fraction (${\varepsilon_{\mathrm{p}}}>0.5$) when compared to analytical results~\citep{Zick1982}.

The diameter of the particle,  which depends on the target solid volume fraction in the box,
determines the total number of Eulerian grid cells at the beginning of the PR-DNS.
Generating a true random assembly of monodispersed spherical particles in a cubic box can be a complex process due to the requirement for a random arrangement with no underlying structure or pattern.
 To achieve this, we have used DEM simulations, by injecting a number of particles in the periodic box with a random velocity, and gradually increase the diameter of the particles until the target solid volume fraction is achieved.
This results in relative particle diameters, $d_\mathrm{p}/L$ ranging from $0.121$  to $0.197$, corresponding to the lowest, 0.1,  and the highest, 0.6, targeted solid volume fraction, respectively. Each PR-DNS simulation contains between 110 and 151 particles, depending on the solid volume fraction.

In the PR-DNS simulations, the flow is driven with 
an external pressure gradient, $\delta P$,
added to the source term of the $x$-direction of the momentum equation, as outlined in Section 2.
The magnitude of the source term varies as a function of the solid volume fraction ${\varepsilon_{\mathrm{p}}}$, and the target mean flow Reynolds number $\text{Re}_\textup{m}$, and is estimated using existing correlations.

A total of six solid volume fractions are studied in this work, ${\varepsilon_{\mathrm{p}}} = 0.1, 0.2, 0.3, 0.4, 0.5,$ and $0.6$.
For each solid volume fraction, multiple realizations are simulated, with flow Reynolds numbers varying from $\text{Re}_\textup{m} = 0.1, 1, 10, 50, 100, $ to $300$.
To obtain converged statistics, three independent realizations of the flow are simulated for each value of mean flow Reynolds number, and solid volume fraction, and each simulation is run until steady state is achieved.

The PR-DNS are performed with a second order spatial scheme for the spatial derivatives,
and a second order temporal scheme~\citep{Denner2020}, with a dynamic time-step based on the convection and viscosity criteria for the temporal derivatives~\citep{Kang2000}.

\section{Results and models for hydrodynamic forces}
\label{sec: Results}

The results of the PR-DNS simulations of the flow past the random assemblies of fixed monodispersed spherical particles are presented in this section, and these results will be used to derive the hydrodynamic force models. To better understand the dynamics of these assemblies, the mean drag force experienced by the particle in the assembly is 
analyzed first. The PR-DNS simulations are compared to existing models presented in the literature, as summarized in table~\ref{tab:IsotropicDragModels}. Based on our results, a new hydrodynamic force model is proposed to predict the mean drag force on a particle in an assembly of monodispersed particles, which varies as a function of the mean flow Reynolds number and the solid volume fraction.\\
\\
In addition to the mean force, the deviation from the mean drag force experienced by the individual particles in the assemblies, due to the local variations in the flow caused by the anisotropy of the particle assembly is also investigated. The local anisotropy is characterized using the scalar, vector, and tensorial Minkowski tensors discussed in Section~\ref{subsec: Minkowski}, which are derived from the Voronoi tessellation of the particle assembly. The Minkowski quantities enable the quantification of the nature of the particle force fluctuations in the assembly. Based on this analysis, dimensionless quantities are defined to characterize the force fluctuations experienced by each particle in the assembly. Using this analysis and the results from the PR-DNS, a novel correlation is derived to predict the forces experienced by each particle in the assembly as a function of these dimensionless quantities.

\subsection{Mean drag force model}
\label{sec: Isotropic result}
\begin{figure}[h!]
	\newcommand{\looselydasheddotted}{\raisebox{2pt}{\tikz{\draw[black,loosely dashdotdotted,line width=1.5pt,mark=x,mark options={scale=1., solid}](0,0) -- (7.5mm,0);}}}
	\newcommand{\dasheddotted}{\raisebox{2pt}{\tikz{\draw[black,dashdotdotted,line width=1.5pt,mark=x,mark options={scale=1., solid}](0,0) -- (7.5mm,0);}}}
	\newcommand{\solidline}{\raisebox{2pt}{\tikz{\draw[black,line width=1.5pt,mark=x,mark options={scale=1., solid}](0,0) -- (7.5mm,0);}}}
	\newcommand{\denselydotted}{\raisebox{2pt}{\tikz{\draw[black,densely dotted,line width=1.5pt,mark=x,mark options={scale=1., solid}](0,0) -- (7.5mm,0);}}}
	\newcommand{\reddashed}{\raisebox{2pt}{\tikz{\draw[red,dashed,line width=1.5pt,mark=x,mark options={scale=1., solid}](0,0) -- (7.5mm,0);}}}
    \sbox1{\looselydasheddotted}\sbox2{\dasheddotted}\sbox3{\solidline}\sbox4{\denselydotted}\sbox5{\reddashed}%
\centering
\includegraphics[width=0.47\textwidth]{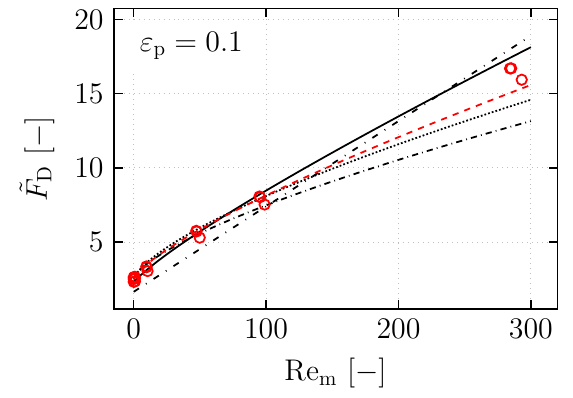}
\includegraphics[width=0.47\textwidth]{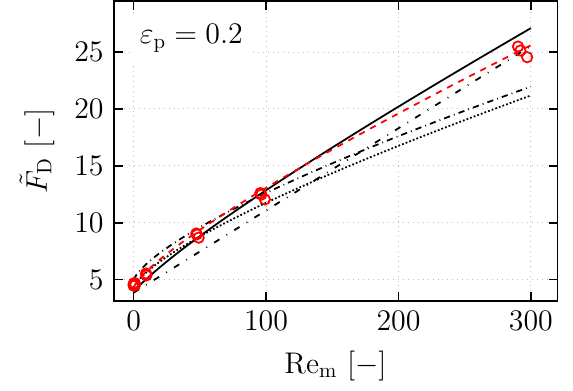}
\includegraphics[width=0.47\textwidth]{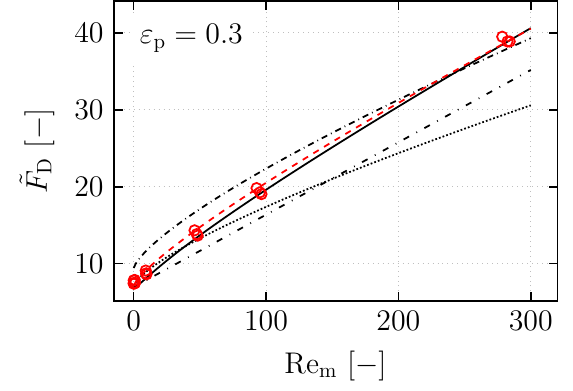}
\includegraphics[width=0.47\textwidth]{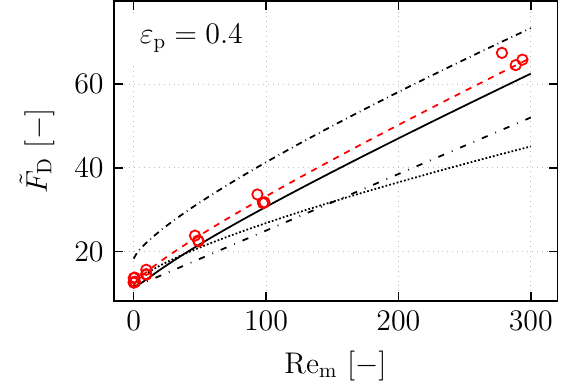}
\includegraphics[width=0.47\textwidth]{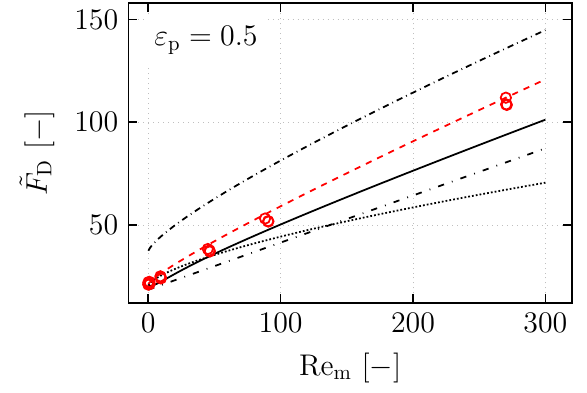}
\includegraphics[width=0.47\textwidth]{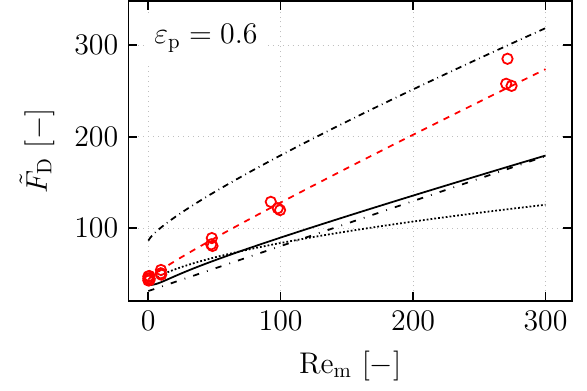}
\caption{Dimensionless average drag force, $\tilde{F}_\mathrm{D}$, as a function of the mean flow Reynolds number, $\text{Re}_\textup{m}$, for the solid volume fraction ${\varepsilon_{\mathrm{p}}} = 0.1, 0.2, 0.3, 0.4, 0.5$ and $0.6$.
Line style indicates correlations predicting the dimensionless mean drag force, 
{\usebox3}:~\citet{Beetstra2007}, {\usebox1}:~\citet{Tenneti2011}, {\usebox4}:~\citet{Tang2015}, {\usebox2}:~\citet{Hill2001b}, {\usebox5}: present correlation: Eq.~\eqref{eq:MeanDragForceCorrelation}, ~\textcolor{red}{$\mathbf{\Circle}$}: present PR-DNS results.}
\label{fig:MeanDragForceResults}
\end{figure}

The mean hydrodynamic force on the particles in the assembly is estimated for each simulation as
\begin{equation}
\avg{\vect{F}_\mathrm{p}} = \frac{1}{N_\mathrm{p}} \sum\limits_{k=1}^{N_\mathrm{p}}\vect{F}_{\mathrm{p},k}
\label{eq:mean drag force}
\end{equation}
where $N_\mathrm{p}$ is the total number of particles in the array, and ${F}_{\mathrm{p},k}$ is the temporally averaged force experienced by the $k$-th particle in the assembly. The mean drag force equals the mean hydrodynamic force in the direction of the mean flow through the particle assembly. In this work, this direction corresponds to the $x$-direction, and therefore the mean drag force is given as $\avg{{F}_{\mathrm{p}x}}$.
The characteristic drag force on a particle in the assembly is obtained by normalizing the mean drag force by the Stokes drag force of a single particle based on the averaged fluid phase velocity to obtain the dimensionless mean drag force, $\tilde{F}_\mathrm{D}$
\begin{equation}
\tilde{F}_\mathrm{D} = \frac{\avg{F_{\mathrm{p}x}}}{3 \pi d_\mathrm{p} \mu \avg{U_x}}
\label{eq:normalized mean drag force}
\end{equation}\\
\\
The dimensionless mean drag force on the particles in the assembly, $\tilde{F}_\mathrm{D}$, as a function of the mean flow Reynolds number, $\text{Re}_\textup{m}$, is shown in figure~\ref{fig:MeanDragForceResults} for all investigated solid volume fractions. 
The reference particle drag correlations given in table~\ref{tab:IsotropicDragModels} are also shown in figure~\ref{fig:MeanDragForceResults}.
In general, our PR-DNS results are in very good agreement with the reference correlations from the literature, up to a solid volume fraction of~${\varepsilon_{\mathrm{p}}} = 0.4$.
Beyond this solid volume fraction, the existing correlations given in table~\ref{tab:IsotropicDragModels} deviate strongly from each other.
For example, at (${\varepsilon_{\mathrm{p}}}$,$\text{Re}_\textup{m}$) = (0.6,300), the correlation of~\citet{Tenneti2011} predicts a mean characteristic drag force of the assembly three times larger than the correlation of~\citet{Tang2015}. This is because most reference correlations are based upon data of particle assemblies with a particle volume fraction up to 0.4.
For all flow regimes, our results lie in between 
the correlations of \citet{Tenneti2011} and \citet{Tang2015}
for the solid volume fractions larger than 0.4. As our PR-DNS framework is especially designed to be accurate at large solid volume fractions~\citep{Cheron2023a}, and we have been able to successfully simulate the flow in assemblies of average solid volume fractions up to 0.6, we believe that the accuracy of existing models for the average drag can be further improved.
Since our data for the lower solid volume fraction is in good agreement with the expression proposed by~\citet{Tenneti2011}, we use this form of the expression, and improve the coefficients using our PR-DNS data. For this novel model to include the correct drag force prediction for an isolated particle, the correlation of~\citet{Schiller1933} is used as a limit of the model as the particle volume fraction approaches zero.
The expression with the newly fitted constants is given by
\begin{align}
\label{eq:MeanDragForceCorrelation}
            \tilde{F}_\mathrm{D}(\varepsilon_{\mathrm{p}}, \mathrm{Re_m}) &= \frac{1 +0.15 \mathrm{Re_m}^{(0.687)}}{(1-\varepsilon_{\mathrm{p}})^{2.20}} + \frac{6.337\varepsilon_{\mathrm{p}}}{(1-\varepsilon_{\mathrm{p}})^{3.}} - \frac{0.652 \varepsilon_{\mathrm{p}}^{1/3}}{(1-\varepsilon_{\mathrm{p}})^4}\\\nonumber &+\varepsilon_{\mathrm{p}}^{0.987} \mathrm{Re_m}\left[ 0.158 + \frac{1.352 \times 10^{-2}}{(1-\varepsilon_{\mathrm{p}})^{4.364}}\right]
\end{align}
This correlation is also shown in figure~\ref{fig:MeanDragForceResults}, and a very good agreement is obtained between the model fit and the DNS data, as the coefficient of determination of our fit is equal to~$\mathcal{R}^{2} = 0.985$.
Large differences with the reference correlation of~\citet{Tenneti2011} are observed for the coefficients multiplied by both the mean flow Reynolds number and the solid volume fraction. Although for low solid volume fraction the original expression of~\citet{Tenneti2011} shows similar results as the correlation proposed in this work, at solid volume fractions larger than 0.4, the newly proposed correlation shows a much better agreement with our PR-DNS data.

\FloatBarrier

\subsection{Deviations from the mean hydrodynamic force}
\label{sec: Mean deviation force}
\subsubsection{Drag force}
\begin{figure}[htbp!]
\centering
\includegraphics[width=0.99\textwidth]{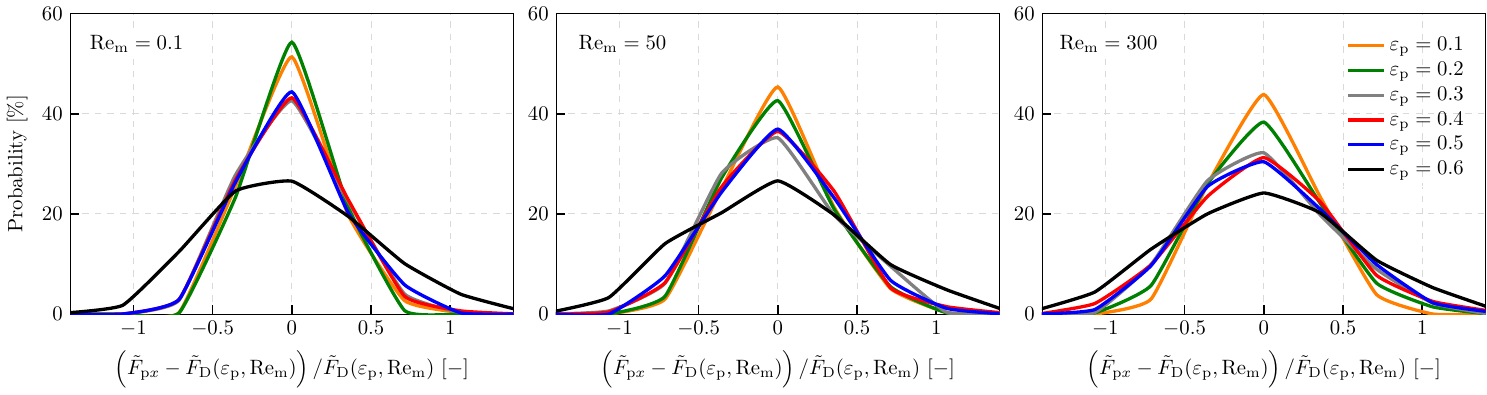}
\caption{PDF of the deviation of the dimensionless drag force per particle from the dimensionless mean drag force of the assembly, scaled by the dimensionless mean drag force of the assembly $\left(\tilde{F}_{\mathrm{p}x} - \tilde{F}_\mathrm{D}(\varepsilon_{\mathrm{p}}, \text{Re}_\textup{m})\right)/\tilde{F}_\mathrm{D}(\varepsilon_{\mathrm{p}}, \text{Re}_\textup{m})$ for all the solid volume fractions and three mean flow Reynolds number $\text{Re}_\textup{m} = 0.1, 50$ and $300$ (from left to right).}
\label{fig:Deviation from the mean colinear force}
\end{figure}

Although the mean drag force correlation provides a very good prediction for the mean drag force of the particles in the assembly, the drag force on the individual particles in the assembly deviates significantly from the average. To show the spread of the deviations in the drag force of the individual particles from the mean drag force in the assembly, the 
histogram of the dimensionless drag force on the individual particles, $\tilde{F}_{\mathrm{p}x}$ scaled by the dimensionless mean drag force of the assembly $\tilde{F}_\mathrm{D}(\varepsilon_{\mathrm{p}}, \text{Re}_\textup{m})$ is shown in
figure~\ref{fig:Deviation from the mean colinear force} for all the solid volume fractions considered and three mean flow Reynolds numbers $\text{Re}_\textup{m} = 0.1, 50$, and $300$.\\
\\
For all mean flow Reynolds numbers, the magnitude of the deviations of the drag force on the individual particles from the mean drag force increases as the solid volume fraction increases.
The mean flow Reynolds number also has an effect on the width of the distribution of the deviation of the drag force.
Especially at the lower solid volume fractions, the width of the distribution increases as the mean flow Reynolds number increases, which is in line with earlier findings~\cite{Akiki2017a,Hardy2022}.
Although this trend is also present for the simulations with a higher solid volume fraction, in these cases this effect is less pronounced.
From these observations, it is clear that the deviations of the drag force on the individual particles from the mean drag force depend on the mean flow Reynolds number as well as the solid volume fraction.

\begin{figure}[htbp!]
\centering
\includegraphics[width=0.59\textwidth]{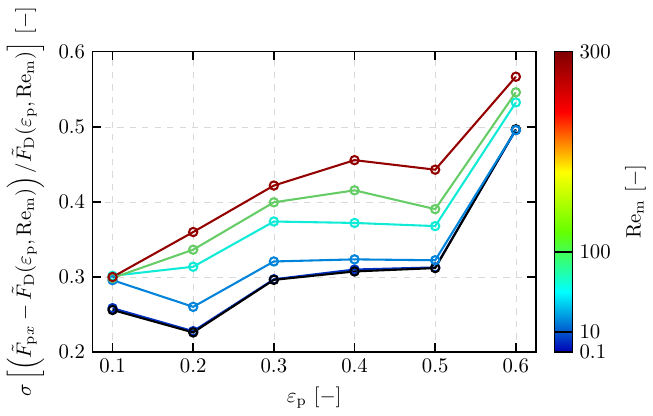}
\caption{
Standard deviation of $\left(\tilde{F}_{\mathrm{p}x} - \tilde{F}_\mathrm{D}(\varepsilon_{\mathrm{p}}, \text{Re}_\textup{m})\right)/\tilde{F}_\mathrm{D}(\varepsilon_{\mathrm{p}}, \text{Re}_\textup{m})$, as a function of the solid volume fraction for all the studied flow regimes $\text{Re}_\textup{m}$ (color bar).}
\label{fig:Evolution of the standard deviation of Fx}
\end{figure}

To provide a quantitative analysis, the standard deviation of distribution of the drag force deviations from the mean drag for all flow regimes and solid volume fractions is shown in figure~\ref{fig:Evolution of the standard deviation of Fx}.
These results confirm that the increase in both the solid volume fraction and the mean flow Reynolds number increase the standard deviation of the distribution of the deviation in the drag forces.\\
\\

\subsubsection{Lift force} \label{sec: Lift force}
\begin{figure}[htbp!]
\centering
\includegraphics[width=0.99\textwidth]{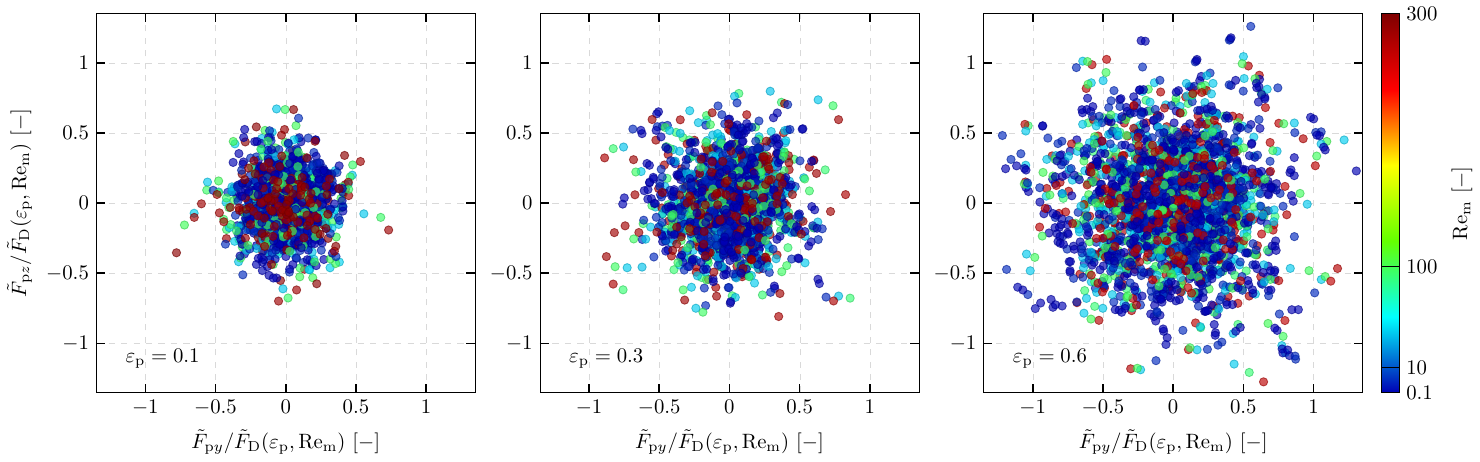}
\caption{Scatter plots of the two Cartesian components of the dimensionless lift force,
$\tilde{F}_{\mathrm{p}y}$ and $\tilde{F}_{\mathrm{p}z}$, for all the mean flow Reynolds number and three solid volume fractions ${\varepsilon_{\mathrm{p}}} = 0.1, 0.3$ and $0.6$ (from left to right).}
\label{fig:ScatterPlotTransverseComponents}
\end{figure}

Although the flow through a uniform assembly of particles does not induce an average force in the direction perpendicular to the direction of the flow, \textit{i.e.} a lift force, the individual particles can experience a force in the direction perpendicular to the flow. This transverse component of the hydrodynamic force can be divided into two Cartesian components. As the mean flow in our simulations is in the $x$-direction, the $y$ and $z$-directions make up the lift force on the individual particles,
\begin{equation}
    \vect{F}_{\mathrm{p}L} = F_{\mathrm{p}y}\vect{e}_y + F_{\mathrm{p}z}\vect{e}_z
\end{equation}
where the direction in which the lift force acts is given by
\begin{equation}
    \dfrac{\vect{F}_{\mathrm{p}L}}{\left| \vect{F}_{\mathrm{p}L} \right|}
     = \cos(\theta)\vect{e}_y + \sin(\theta)\vect{e}_z
\end{equation}
with the magnitude of the lift force being
\begin{equation}
    \left| \vect{F}_{\mathrm{p}L} \right| = \sqrt{F_{\mathrm{p}y}^2 + F_{\mathrm{p}z}^2}
\end{equation}
There are two options to create a model for the lift force on the individual particles in the assembly. The first is to have a model to predict the magnitude and the angle of the lift force, and the second option is a model to predict the two Cartesian components of the lift force. In this work, we model the lift forces using the latter approach, \textit{i.e.} the two Cartesian components perpendicular to the direction of the flow.
The advantage of this approach is that the statistical properties of the forces in the two Cartesian directions, $y$ and $z$, are 
identical for each simulation case. Therefore, gathering the statistics of one of the Cartesian components suffices. 
This is confirmed by analyzing the two Cartesian components of the lift force, given in dimensionless form by dividing each force component by the mean drag force on a particle (Eq.~\eqref{eq:normalized mean drag force}), plotted in figure~\ref{fig:ScatterPlotTransverseComponents} for three of the solid volume fractions, and we found their statistics (mean and standard deviation) to be equal.
Therefore, we only focus on a single component of the lift force.\\
\\

\begin{figure}[htbp!]
\centering
\includegraphics[width=0.99\textwidth]{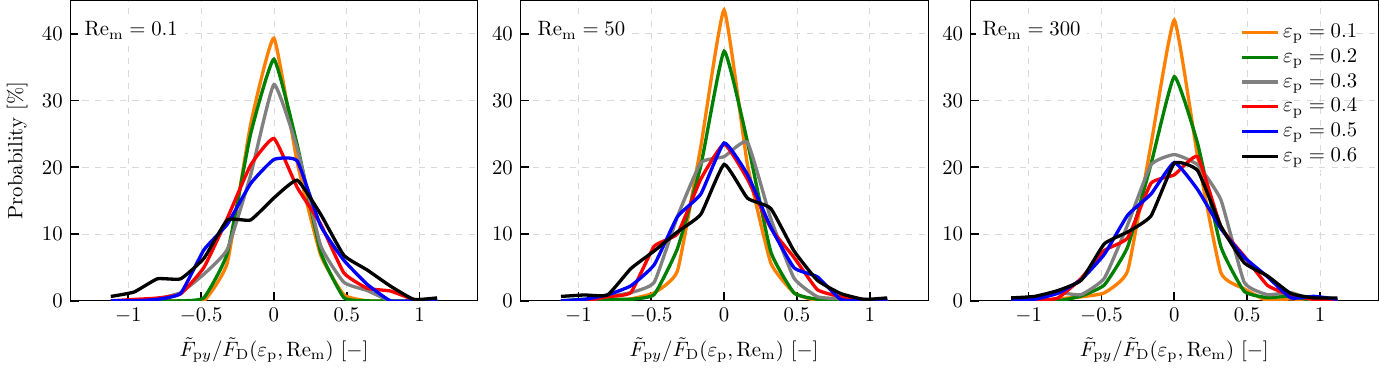}
\caption{Histogram of the dimensionless lift force per particle scaled by the dimensionless mean drag force of the assembly $F_{\mathrm{p}y}/\tilde{F}_\mathrm{D}(\varepsilon_{\mathrm{p}}, \text{Re}_\textup{m})$ for all the solid volume fractions and three mean flow Reynolds number $\text{Re}_\textup{m} = 0.1, 50$ and $300$ (from left to right).}
\label{fig:Deviation transverse force from the mean colinear force}
\end{figure}

To examine the variations in the lift force component, the PDF of the lift forces scaled by the mean drag force on the particle assembly are shown in figure~\ref{fig:Deviation transverse force from the mean colinear force} for all the solid volume fractions and three mean flow Reynolds numbers considered in this work. Similar to the findings for the deviations in the drag force compared to the mean drag force of the assembly, the spread of the distribution of the lift force increases with increasing solid volume fraction, regardless of the mean flow Reynolds number.
In contrast to the spread in the deviation of the drag force, the spread in the lift force does not significantly depend on the mean flow Reynolds number.
Therefore, predicting the maximum deviation in the lift force depends primarily on the solid volume fraction and only weakly depends on the mean flow Reynolds number.\\
\\

\begin{figure}[htbp!]
\centering
\includegraphics[width=0.99\textwidth]{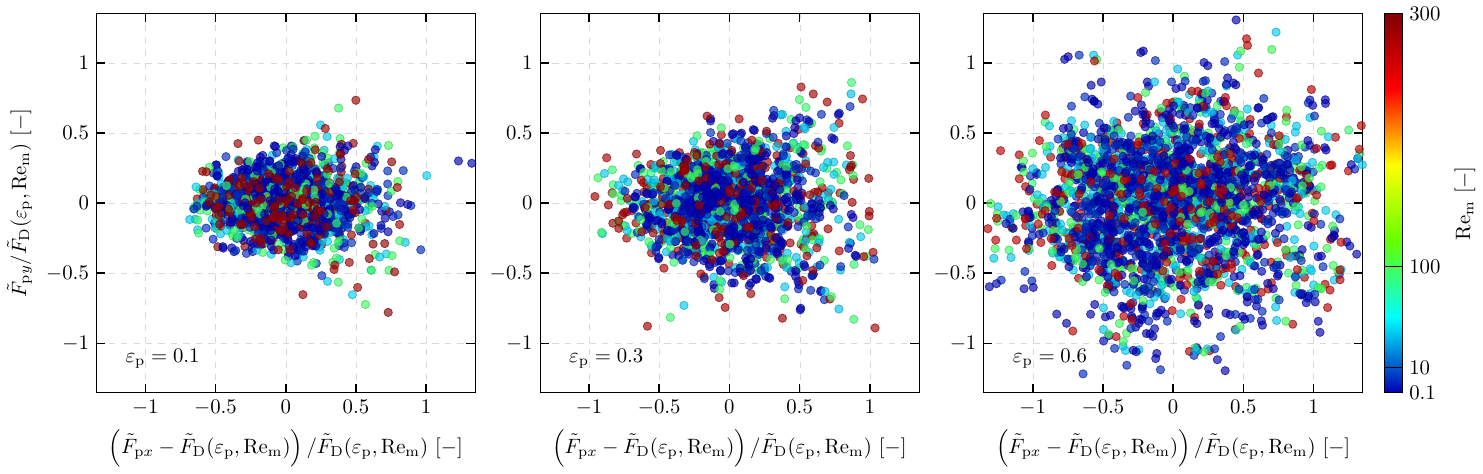}
\caption{Scatter plots between the lift and drag forces components, $\tilde{F}_{\mathrm{p}x}$ and $\tilde{F}_{\mathrm{p}y}$, scaled by the mean drag force of the assembly $\tilde{F}_\mathrm{D}(\varepsilon_{\mathrm{p}}, \text{Re}_\textup{m})$, for all the mean flow Reynolds number and three solid volume fractions ${\varepsilon_{\mathrm{p}}} = 0.1, 0.3$ and $0.6$ (from left to right).}
\label{fig:ScatterPlotPerpParallelComponents}
\end{figure}

In some configurations concerning objects in flows, the lift and drag forces are related~\citep{Brenner1963}, and the existence of this possible relation is also investigated for the case of particles in an assembly. Figure~\ref{fig:ScatterPlotPerpParallelComponents} shows the scatter plots of the drag force deviations from the mean drag force versus one component of the lift force for three solid volume fractions, for all mean flow Reynolds numbers for each solid volume fraction.
For the solid volume fractions ${\varepsilon_{\mathrm{p}}} = 0.1$ and $0.3$ there is a slight increase in the spread of the lift force when the spread in the drag force increases. This trend is somewhat more pronounced as the mean flow Reynolds number increases.
However, this behavior is not observed for all solid volume fractions, and the lift force deviations are not correlated with the deviations of the drag force. Therefore, in this work we have chosen to create independent models; one for the drag force deviations and another for the lift forces on the individual particles, as there seems not to be sufficient correlation between the two to warrant success of a single model.

\subsection{Correlations between the Voronoi characterization and the forces in the assemblies}
\label{sec: Voronoi analysis of deviation}
All the variables introduced in this work to describe the microstructure based on the Minkowski tensors as outlined in section~\ref{subsec: Minkowski} constitute a very high dimensional parameter space. Aiming for the final expressions predicting the hydrodynamic drag forces to contain all these variables is cumbersome and also unnecessary.
In order to reduce the large multidimensional parameter space, singular value decomposition is used to explore the possible dependencies between the characterization of the assembly, the flow conditions, and the resulting hydrodynamic forces. This analysis allows to identify and retain only the variables that effectively capture the correlations between the characterization of the microstructure, the flow conditions, and the hydrodynamic forces.
The correlation between the variables characterizing the microstructure and the hydrodynamic forces is examined by computing the Pearson correlation coefficient between the two independent variables $X$ and $Y$, as
\begin{equation}
\mathcal{R} = \left\vert\left\vert\frac{\mathrm{Cov}\avg{X Y}}{\sigma(X) \sigma(Y)}\right\vert\right\vert
\end{equation}
where $\sigma(X)$ is the standard deviation of the variable $X$, and $\mathrm{Cov}$ is the covariance between the two independent variables given by
\begin{equation}
\mathrm{Cov}\avg{X Y} = \avg{X^{'} Y^{'}}
\end{equation}
where $X^{'}$ is the deviation of the variable $X$ with respect to its average $\avg{X}$. 
The Pearson correlation coefficient indicates the extent of correlation between two variables.
This Pearson correlation coefficient lies between $0 \leq \mathcal{R} \leq 1$, where $\mathcal{R} = 0$ indicates no correlation and $\mathcal{R} = 1$ indicates a perfect correlation between both variables. This coefficient is used in our work to assess the quality of a model fit prediction,
and to assess if a variable should be retained or rejected when creating a correlation.

\subsubsection{Correlation between the assembly structure and the drag force deviations} 
\label{sec: Voronoi analysis colinear force}

Table~\ref{tab:Pearson Coefficient Colinear Force} shows the Pearson correlation coefficient 
of the variables taken from the Minkowski tensors which correlate most to the drag force deviations in the particle assembly.
The local solid volume fraction is also shown in table~\ref{tab:Pearson Coefficient Colinear Force}.
The Pearson correlation coefficient is determined for each examined solid volume fraction separately, as well as for the entire set of data.

\begin{table}[h]
    \begin{tabular}{l|c|c|ccc|ccc}
        \textbf{} & $\varepsilon_{\mathrm{V}}$ & $\left(\lambda^{\textup{2,0}}_{\textup{1}}\right)_{y}$ & $\left(\varepsilon_{W^{\textup{2,0}}_{\textup{0}}}\right)_{x,x}$ & $\left(\varepsilon_{W^{\textup{2,0}}_{\textup{0}}}\right)_{y,y}$  & $\left(\varepsilon_{W^{\textup{2,0}}_{\textup{0}}}\right)_{z,z}$ & $\left(\varepsilon_{W^{\textup{0,2}}_{\textup{1}}}\right)_{x,x}$ & $\left(\varepsilon_{W^{\textup{0,2}}_{\textup{1}}}\right)_{y,y}$ & $\left(\varepsilon_{W^{\textup{0,2}}_{\textup{1}}}\right)_{z,z}$\\
    \hline
    \hline
    $\mathcal{R}\left({\varepsilon_\mathrm{p}} = 0.1\right)$ & 0.268 & 0.218 & 0.443& 0.174 &0.233  &0.525 &0.219 &0.306 \\[1mm]
    $\mathcal{R}\left({\varepsilon_\mathrm{p}} = 0.2\right)$ & 0.082 & 0.255 & 0.593& 0.241 &0.308  &0.632 & 0.258 &0.369  \\[1mm]
    $\mathcal{R}\left({\varepsilon_\mathrm{p}} = 0.3\right)$ & 0.143 & 0.375 & 0.557& 0.373 &0.356  &0.627 &0.364 &0.343 \\[1mm]
    $\mathcal{R}\left({\varepsilon_\mathrm{p}} = 0.4\right)$ & 0.169 & 0.243 & 0.463& 0.324 &0.393  &0.636 &0.266 &0.368 \\[1mm]
    $\mathcal{R}\left({\varepsilon_\mathrm{p}} = 0.5\right)$ & 0.144 & 0.291 & 0.392& 0.306 &0.310  &0.587 &0.311 &0.293 \\[1mm]
    $\mathcal{R}\left({\varepsilon_\mathrm{p}} = 0.6\right)$ & 0.355 & 0.37 & 0.193& 0.466 &0.432  &0.560 &0.303 &0.257 \\[1mm]
    \hline
    $\mathcal{R}\left(\sum {\varepsilon_\mathrm{p}}\right)$ & 0.01 & 0.245 & 0.348& 0.210 &0.232  &0.499 &0.234 &0.273 \\
    \end{tabular}
    \caption{The Pearson correlation coefficient, $\mathcal{R}$, between the dimensionless drag force for each of the volume fractions considered in this work, and the combined data set, and the various components from the Minkowski tensors.}
    \label{tab:Pearson Coefficient Colinear Force}
\end{table}

\begin{figure}[htbp!]
    \centering
    \includegraphics[width=0.65\textwidth]{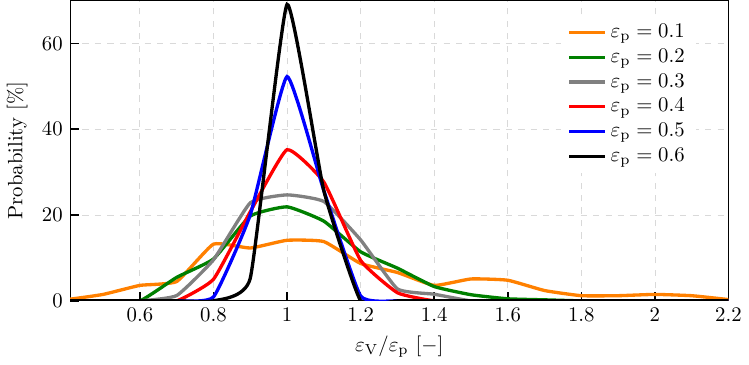}
    \caption{Histogram of the relative local solid volume fraction, $\varepsilon_\mathrm{V}$ for each of the averaged solid volume fractions, $\varepsilon_\mathrm{p}$.}
    \label{fig:HistogramScaledVolumefraction}
    \end{figure}

\begin{figure}[htbp!]
    \centering
    \includegraphics[width=0.99\textwidth]{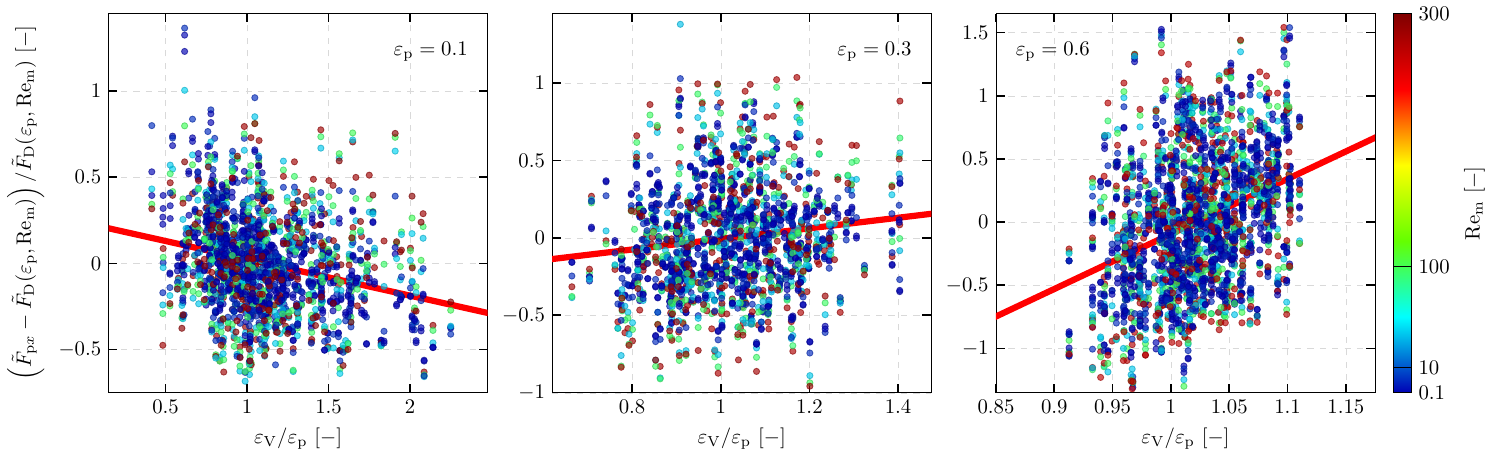}
    \includegraphics[width=0.99\textwidth]{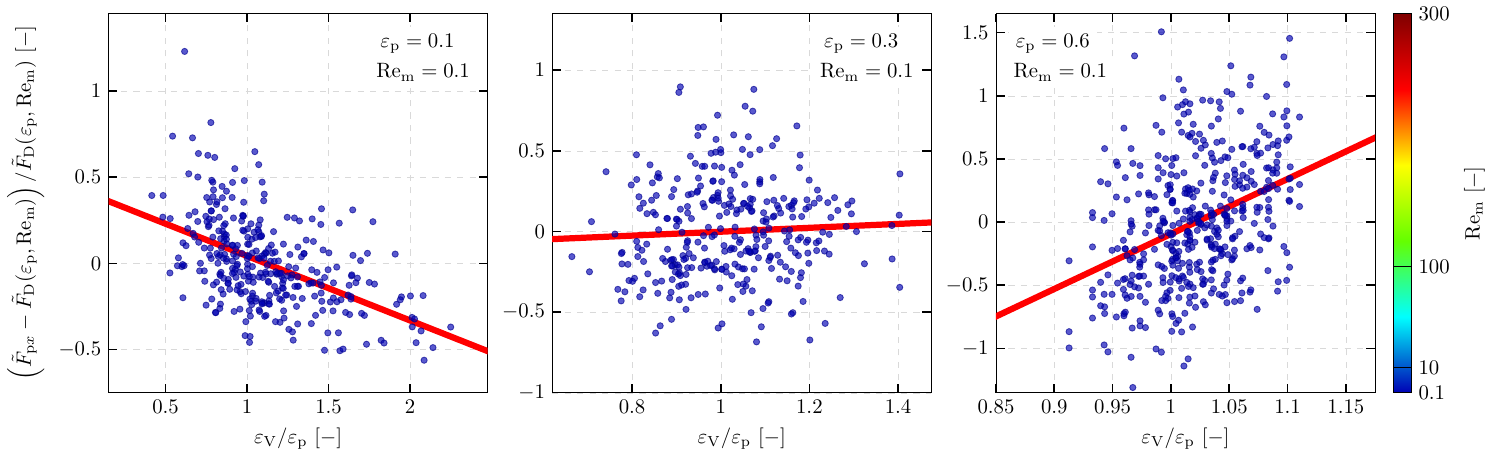}
    \includegraphics[width=0.99\textwidth]{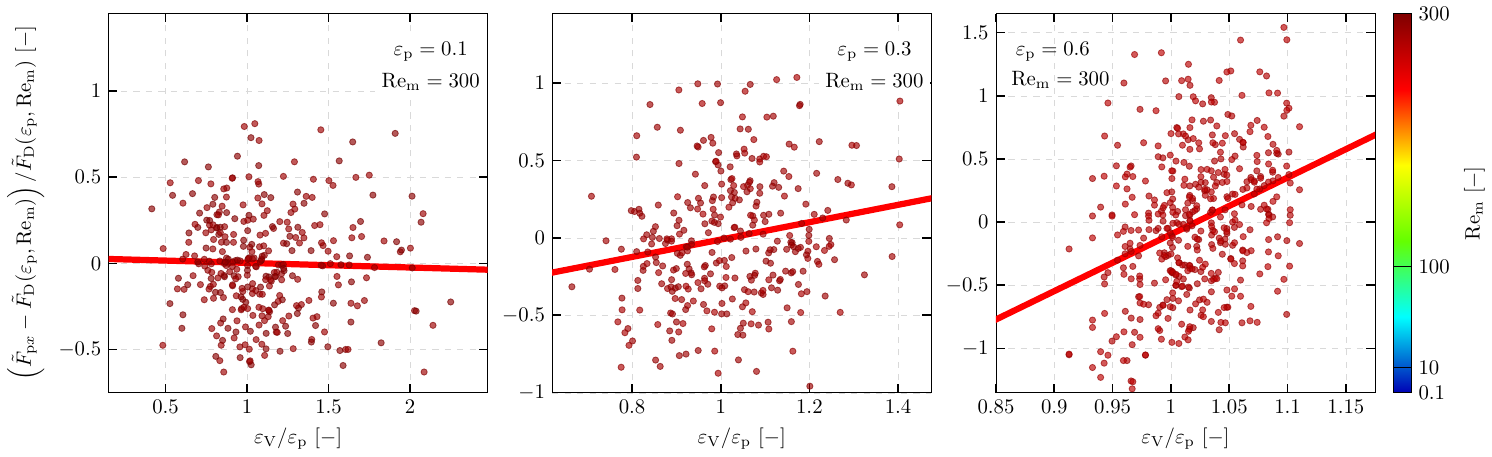}
    \caption{Scatter plots of the relative drag force of the individual particles as a function of the relative local solid volume fraction. The top row shows the relative drag force for all mean Reynolds numbers for 3 different mean solid volume fractions, the middle row shows the relative drag force for the mean Reynolds number of 0.1  for 3 different mean solid volume fractions, and the bottom row shows the relative drag force for the mean Reynolds number of 300  for 3 different mean solid volume fractions. The red line indicates the regression trend between the variables.}
    \label{fig:ScatterPlotFParallelLocalVolumeFraction}
\end{figure}

Figure~\ref{fig:HistogramScaledVolumefraction} shows the histogram of the occurrence of the relative local volume fraction. As expected, the denser the particle assembly, the narrower the distribution. 
However, the local solid volume fraction does not generally correlate with the deviation of the drag force for the entire dataset, which is an unexpected result.  Only for ${\varepsilon_\mathrm{p}} = 0.1$ and $0.6$ a significant correlation is found, for the other volume fractions the correlation is weak.
To investigate this further, the individual relative drag forces on the particles in the assembly are shown for three solid volume fractions in Figure~\ref{fig:ScatterPlotFParallelLocalVolumeFraction}. In the top three figures, the collated data for all mean Reynolds numbers is shown, in the middle row the relative drag force on the particles for the lowest mean Reynolds number is shown, and in the bottom three figures the relative drag for the highest mean Reynolds numbers is shown. In all the figures, the trend line is shown in red.

From the top left plot in figure~\ref{fig:ScatterPlotFParallelLocalVolumeFraction}, showing the results for the lowest mean volume fraction, it can be seen that the relative drag force on a particle generally decreases as the local volume fraction increases. However, for the highest volume fraction, shown in the third plot on the top row, this trend is clearly reversed, and the relative drag force on the individual particles increases as the local volume fraction increases.

There are two potential effects of the local solid volume fraction. The first effect is, that with increasing local solid volume fraction, the interstitial fluid flow increases, thereby increasing the drag force. In the second effect, areas of relative low local solid volume fraction are `voids' in the particle assembly, and consequently have a lower relative fluid pressure, which increases the actual fluid velocity into that void. This also increases the drag force. It seems that in the cases with the low mean solid volume fraction the first effect dominates, and for high mean solid volume fraction, the second effect dominates. For the volume fractions between 0.1 and 0.6, shown by the middle column of plots in Figure~\ref{fig:ScatterPlotFParallelLocalVolumeFraction}, these two effects more or less balance, leading to a very weak correlation between the local solid volume fraction and the relative drag force on the individual particles.

\begin{figure}[htbp!]
\centering
\includegraphics[width=0.99\textwidth]{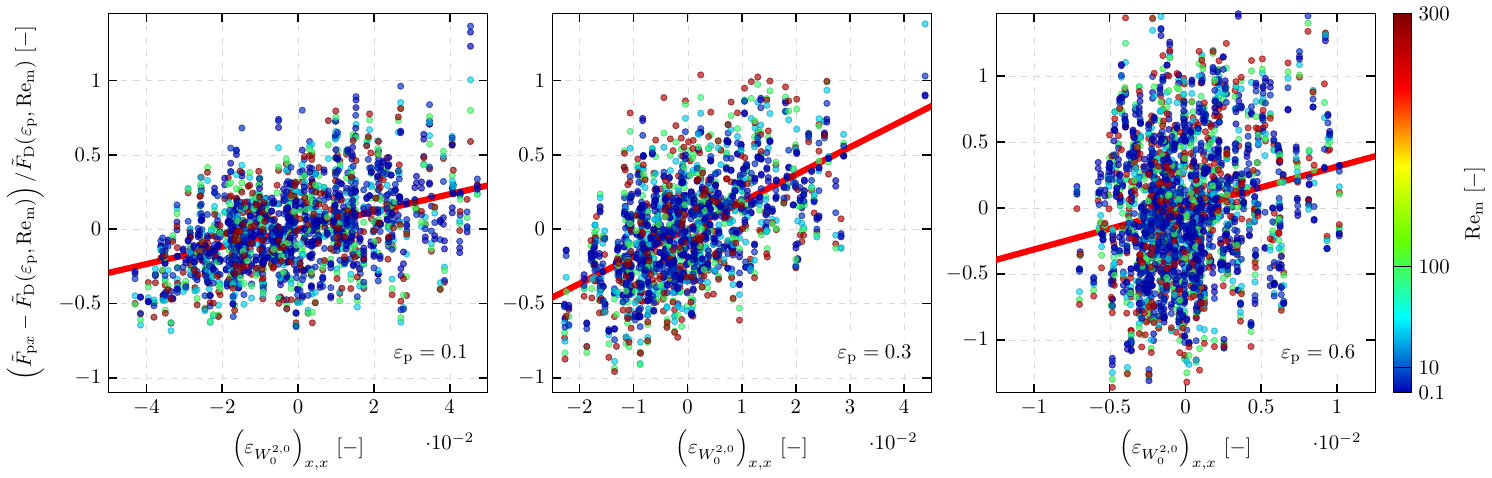}
\includegraphics[width=0.99\textwidth]{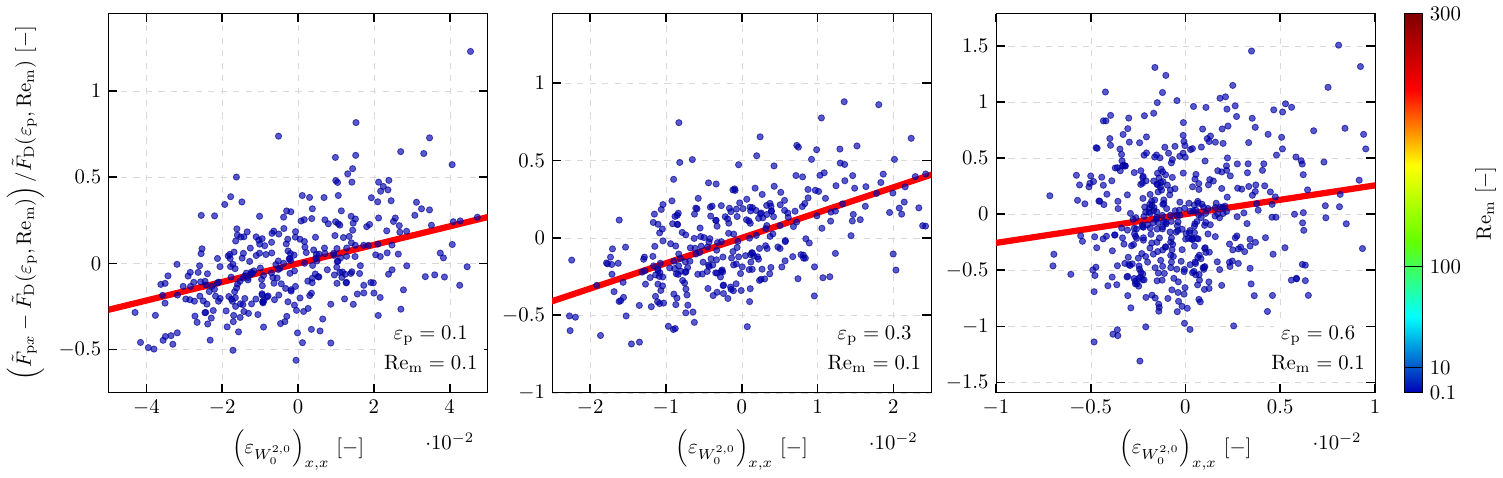}
\includegraphics[width=0.99\textwidth]{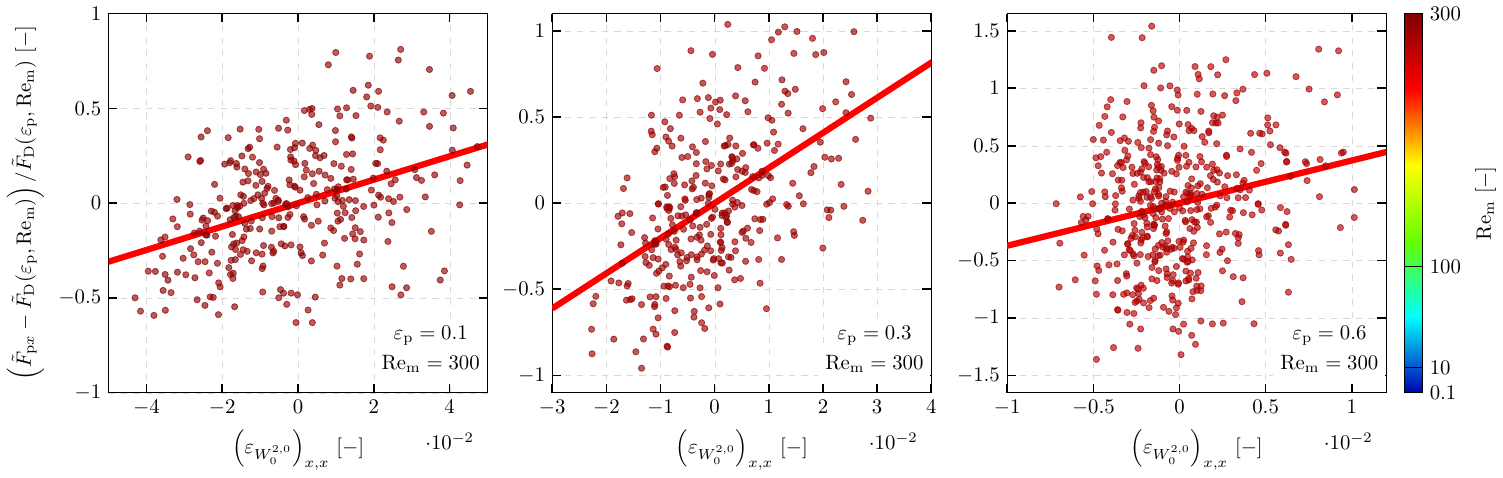}
\caption{
Scatter plots of the relative drag force of the individual particles as a function of the $x,x$ component of the Minkowski tensor $\tens{\varepsilon}_{W^{\textup{2,0}}_{\textup{0}}}$. The top row shows the relative drag force for all mean Reynolds numbers for 3 different mean solid volume fractions, the middle row shows the relative drag force for the mean Reynolds number of 0.1  for 3 different mean solid volume fractions, and the bottom row shows the relative drag force for the mean Reynolds number of 300  for 3 different mean solid volume fractions. The red line indicates the regression trend between the variables.}    
\label{fig:ScatterPlotFParallel Voro-WTwoZeroZerotensor-00}
\end{figure}

The other variables derived from the Minkowski tensors that best correlate with the variation in the relative drag forces on the individual particles are shown in table~\ref{tab:Pearson Coefficient Colinear Force}.
These variables are the $x,x$ component of the two tensors $\tens{\varepsilon}_{W^{\textup{2,0}}_{\textup{0}}}$ and $\tens{\varepsilon}_{W^{\textup{0,2}}_{\textup{1}}}$. Figure~\ref{fig:ScatterPlotFParallel Voro-WTwoZeroZerotensor-00} shows the relative drag force on the individual particles in the assembly as a function of the $x,x$ component of the tensor $\tens{\varepsilon}_{W^{\textup{2,0}}_{\textup{0}}}$; the top row showing the results for all mean Reynolds numbers, the middle row showing the results of the lowest mean Reynolds number, and the bottom row showing the results for the highest mean Reynolds number. The trends for all the configurations studied in this paper are the same; as the value for the $x,x$ component of the tensor $\tens{\varepsilon}_{W^{\textup{2,0}}_{\textup{0}}}$ increases, the magnitude of the relative drag on the individual particle in the assembly increases. Moreover, it can be seen from the bottom two rows of figure~\ref{fig:ScatterPlotFParallel Voro-WTwoZeroZerotensor-00}, that the mean Reynolds number also has an affect. As the mean Reynolds number increases, the trend between the relative drag on the individual particle in the assembly and the $x,x$ component of the tensor $\tens{\varepsilon}_{W^{\textup{2,0}}_{\textup{0}}}$ becomes more pronounced.

\FloatBarrier

\begin{figure}[htbp!]
\centering
\includegraphics[width=0.99\textwidth]{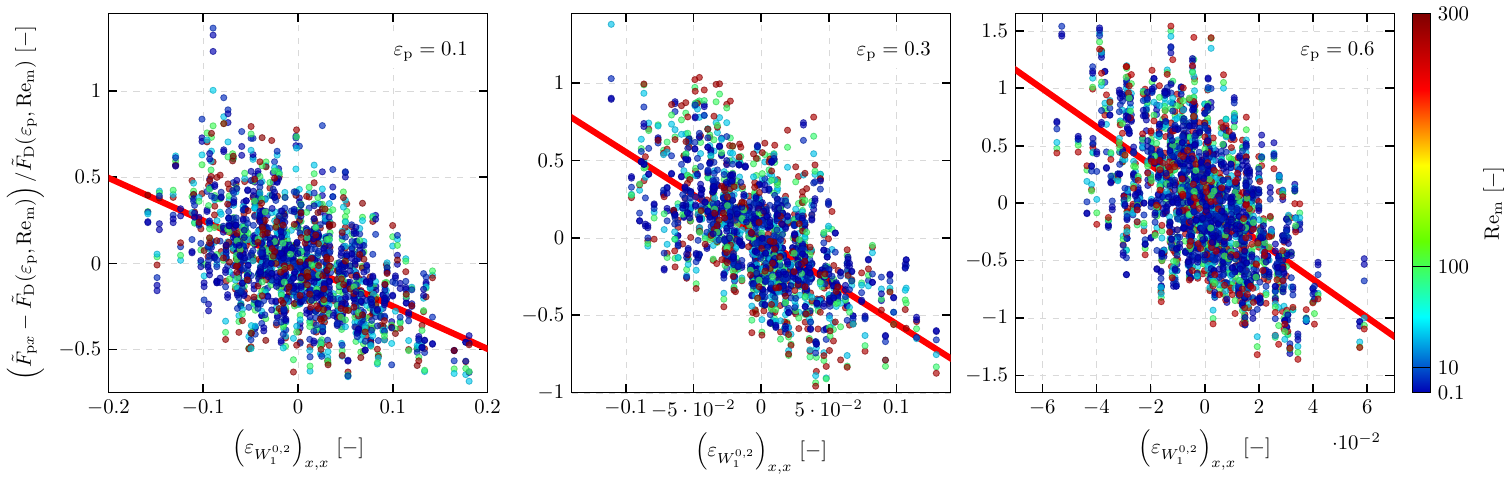}
\includegraphics[width=0.99\textwidth]{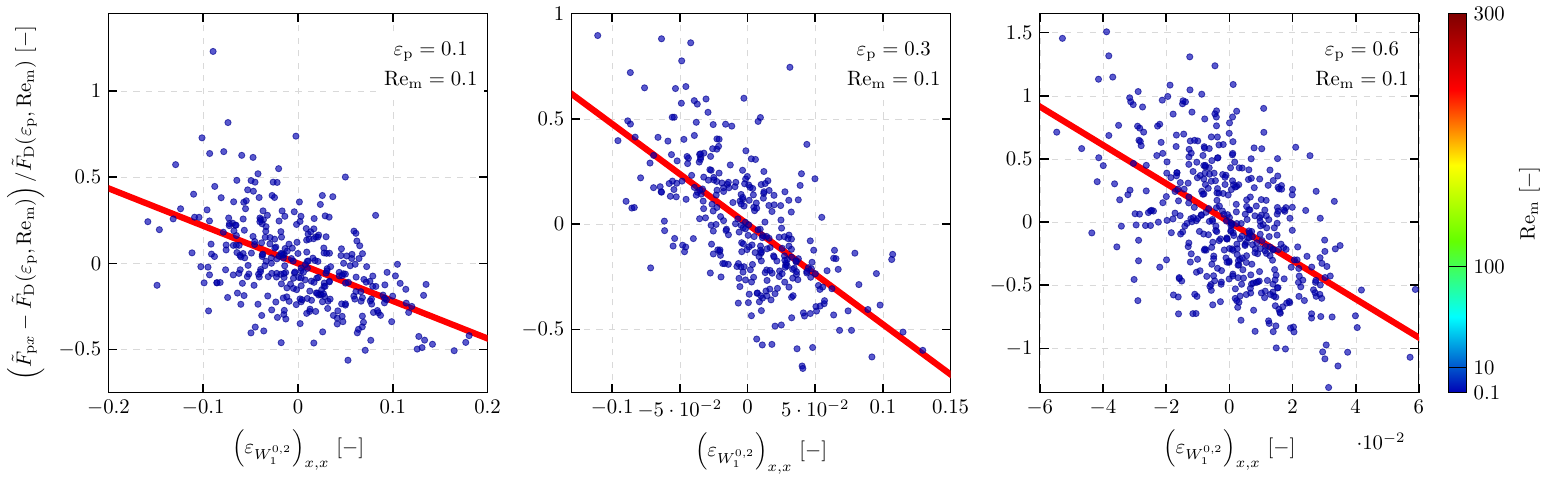}
\includegraphics[width=0.99\textwidth]{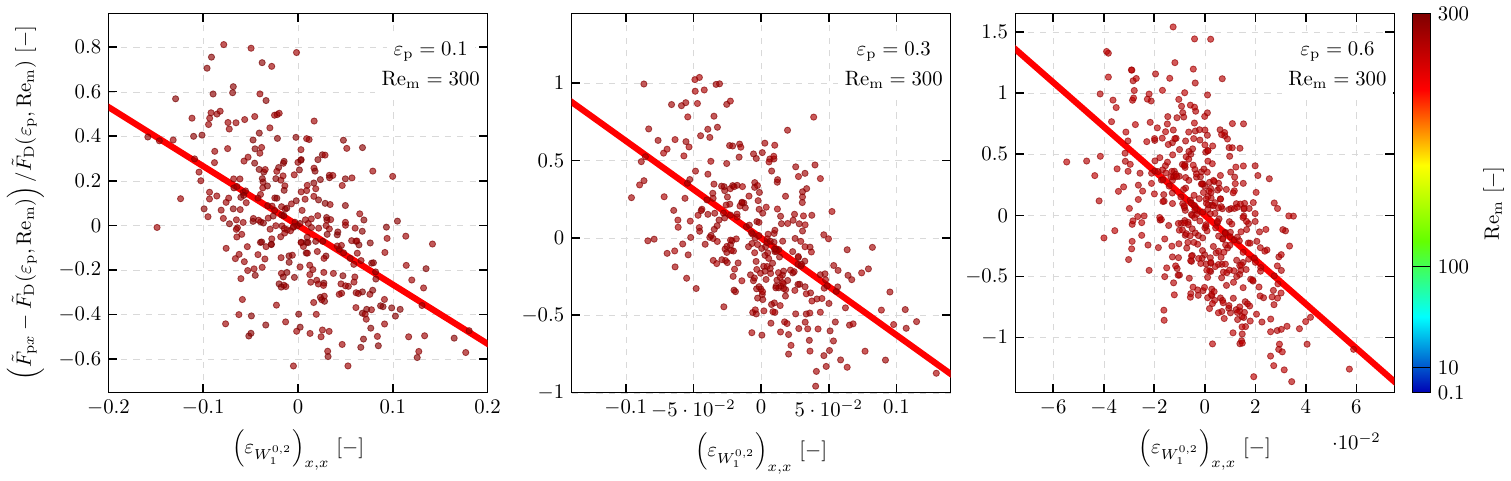}
\caption{Scatter plots of the relative drag force of the individual particles as a function of the $x,x$ component of the Minkowski tensor $\tens{\varepsilon}_{W^{\textup{2,0}}_{\textup{1}}}$. The top row shows the relative drag force for all mean Reynolds numbers for 3 different mean solid volume fractions, the middle row shows the relative drag force for the mean Reynolds number of 0.1  for 3 different mean solid volume fractions, and the bottom row shows the relative drag force for the mean Reynolds number of 300  for 3 different mean solid volume fractions. The red line indicates the regression trend between the variables.}   
\label{fig:ScatterPlotFParallel Voro-WZeroTwoOnetensor-01}
\end{figure}

Figure~\ref{fig:ScatterPlotFParallel Voro-WZeroTwoOnetensor-01} shows the relative drag force on the individual particles in the assembly as a function of the $x,x$ component of the tensor $\tens{\varepsilon}_{W^{\textup{2,0}}_{\textup{1}}}$; the top row showing the results for all mean Reynolds numbers, the middle row showing the results of the lowest mean Reynolds number, and the bottom row showing the results for the highest mean Reynolds number. The correlations for all the configurations studied in this paper are quite strong; as the value for the $x,x$ component of the tensor $\tens{\varepsilon}_{W^{\textup{2,0}}_{\textup{1}}}$ increases, the magnitude of the relative drag on the individual particle in the assembly decreases strongly. This holds to be true for all mean flow Reynolds numbers.

\FloatBarrier

\subsubsection{Correlation between the assembly structure and the lift forces on the particles}
\label{sec: Voronoi analysis transverse force}

In tables~\ref{tab:Pearson Coefficient Transverse Force Y} and~\ref{tab:Pearson Coefficient Transverse Force Z}, the Pearson correlation coefficient, $\mathcal{R}$, of the variables based on the Minkowski tensors, which are most correlated with the lift force components are shown. The correlation is shown for the 6 different mean solid volume fractions, as well as for the complete data set encompassing all mean solid volume fractions.

For most of the variables based on the Minkowski tensors, the largest correlation coefficient is with the component in the same direction as the force component. 
For instance, the most correlated coefficient of the stretching vector $\vect{\mathcal{D}}$ are the $y$ and $z$ components for the force components $F_{\mathrm{p}y}$ and $F_{\mathrm{p}z}$, respectively. This is similar for the components of the $\tens{\varepsilon}_{W^{\textup{2,0}}_{\textup{0}}}$ and $\tens{\varepsilon}_{W^{\textup{0,2}}_{\textup{1}}}$ tensors.
In addition, the Pearson coefficients for the entire data set of the most correlated variables between the lift force components and variables derived from the Minkowski tensors are almost equal, which corroborates the similarities between the two components of the lift force. Therefore, we will only consider the $y$-direction of the lift force, to represent the complete lift force itself, and correlate this with the $y$ components of the variables derived from the Minkowski tensors. To obtain a prediction for the lift force in the $z$-direction, the same relation as the one of the $y$ component can be used, substituting the $z$ components of the variables from the Minkowski tensors.

\begin{table}[h]
    \centering
    \begin{tabular}{l|c|c|c|c|c|c|c}
        \textbf{} & $\varepsilon_{\mathrm{V}}$ & 
        $\mathcal{D}_y$ &
        $\left(\varepsilon_{W^{\textup{1,0}}_{\textup{0}}}\right)_{y}$ &
        $\left(\lambda^{\textup{2,0}}_{\textup{1}}\right)_{y}$ &
        $\left(\varepsilon_{W^{\textup{2,0}}_{\textup{0}}}\right)_{x,y}$ & $\left(\varepsilon_{W^{\textup{0,2}}_{\textup{1}}}\right)_{x,y}$\\
    \hline
    \hline
    $\mathcal{R}\left({\varepsilon_\mathrm{p}} = 0.1\right)$ & 0.051 & 0.374 & 0.298& 0.248 &0.473  &0.521 \\[1mm]
    $\mathcal{R}\left({\varepsilon_\mathrm{p}} = 0.2\right)$ & 0.062 & 0.347 & 0.286& 0.221 &0.571  &0.582 \\[1mm]
    $\mathcal{R}\left({\varepsilon_\mathrm{p}} = 0.3\right)$ & 0.052 & 0.199 & 0.199& 0.146 &0.581  &0.572 \\[1mm]
    $\mathcal{R}\left({\varepsilon_\mathrm{p}} = 0.4\right)$ & 0.032 & 0.213 & 0.238& 0.31 &0.564  &0.562 \\[1mm]
    $\mathcal{R}\left({\varepsilon_\mathrm{p}} = 0.5\right)$ & 0.071 & 0.24 & 0.21& 0.316 &0.601  &0.589 \\[1mm]
    $\mathcal{R}\left({\varepsilon_\mathrm{p}} = 0.6\right)$ & 0.01 & 0.24 & 0.191& 0.187 &0.573  &0.544 \\[1mm]
    \hline
    $\mathcal{R}\left(\sum {\varepsilon_\mathrm{p}}\right)$ & 0.008 & 0.189 & 0.157& 0.192 &0.361  &0.44 \\
    \end{tabular}
    \caption{Pearson correlation coefficient, $\mathcal{R}$, between the dimensionless y-component of the lift force $F_{\mathrm{p}y}$, and the various components of the Minkowski tensors.}
    \label{tab:Pearson Coefficient Transverse Force Y}
\end{table}

\begin{table}[h]
    \centering
    \begin{tabular}{l|c|c|c|c|c|c|c}
        \textbf{} & $\varepsilon_{\mathrm{V}}$ & 
        $\mathcal{D}_z$ &
        $\left(\varepsilon_{W^{\textup{1,0}}_{\textup{0}}}\right)_{z}$ &
        $\left(\lambda^{\textup{2,0}}_{\textup{1}}\right)_{2}$ &
        $\left(\varepsilon_{W^{\textup{2,0}}_{\textup{0}}}\right)_{x,z}$ & $\left(\varepsilon_{W^{\textup{0,2}}_{\textup{1}}}\right)_{x,z}$\\
    \hline
    \hline
    $\mathcal{R}\left({\varepsilon_\mathrm{p}} = 0.1\right)$ & 0.056 & 0.342 & 0.318& 0.142 &0.433  &0.518 \\[1mm]
    $\mathcal{R}\left({\varepsilon_\mathrm{p}} = 0.2\right)$ & 0.068 & 0.446 & 0.415& 0.274 &0.511  &0.535 \\[1mm]
    $\mathcal{R}\left({\varepsilon_\mathrm{p}} = 0.3\right)$ & 0.005 & 0.247 & 0.243& 0.208 &0.646  &0.639 \\[1mm]
    $\mathcal{R}\left({\varepsilon_\mathrm{p}} = 0.4\right)$ & 0.006 & 0.261 & 0.251& 0.2 &0.564  &0.56 \\[1mm]
    $\mathcal{R}\left({\varepsilon_\mathrm{p}} = 0.5\right)$ & 0.12 & 0.182 & 0.171& 0.202 &0.515  &0.495 \\[1mm]
    $\mathcal{R}\left({\varepsilon_\mathrm{p}} = 0.6\right)$ & 0.068 & 0.114 & 0.115& 0.143 &0.584  &0.561 \\[1mm]
    \hline
    $\mathcal{R}\left(\sum {\varepsilon_\mathrm{p}}\right)$ & 0.006 & 0.213 & 0.188& 0.162 &0.36  &0.446 \\
    \end{tabular}
    \caption{Pearson correlation coefficient, $\mathcal{R}$, between the dimensionless z-component of the lift force $F_{\mathrm{p}z}$, and the various components of the Minkowski tensors.}
    \label{tab:Pearson Coefficient Transverse Force Z}
\end{table}

\begin{figure}[htbp!]
\centering
\includegraphics[width=0.99\textwidth]{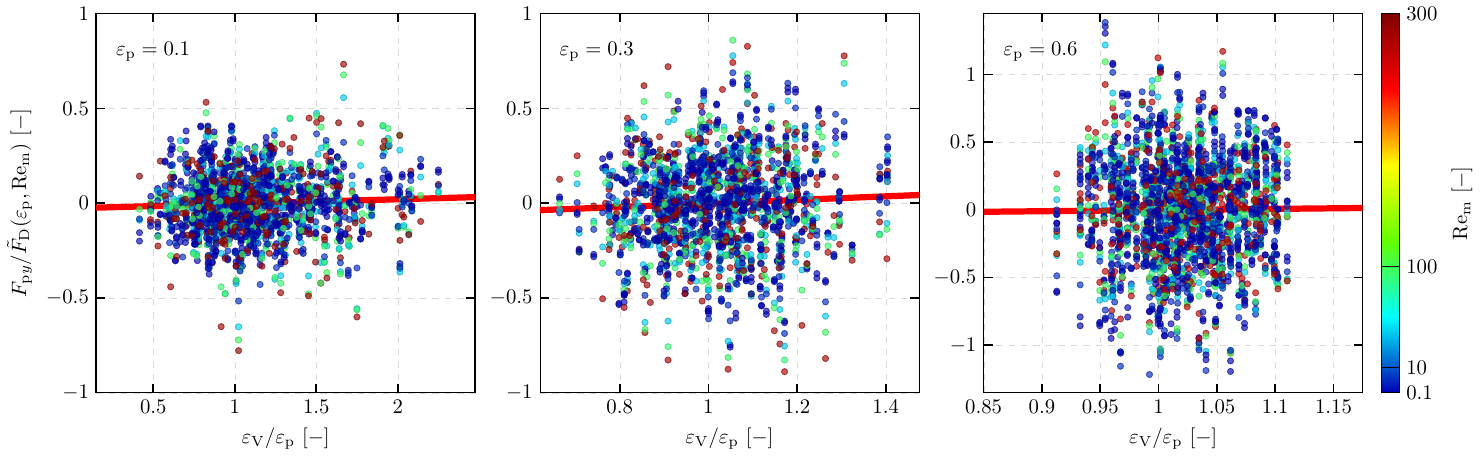}
\caption{Scatter plots between the $y$ component of the lift force scaled by the mean drag force on the individual particles in the assembly for three different mean solid volume fractions, 0.1, 0.3 and 0.6. The red line shows the general trend. The red line indicates the regression trend between the variables.}
\label{fig:ScatterPlotFLiftParallelLocalVolumeFraction}
\end{figure}

As for the relative drag force, the correlation between the local solid volume fraction and the lift forces on the individual particles is unexpectedly low. 
The scatter plots of the relative lift forces as a function of the local solid volume fraction for three mean solid volume fractions are shown in 
figure~\ref{fig:ScatterPlotFLiftParallelLocalVolumeFraction}, and there is no clear observable trend.

\begin{figure}[htbp!]
\centering
\includegraphics[width=0.99\textwidth]{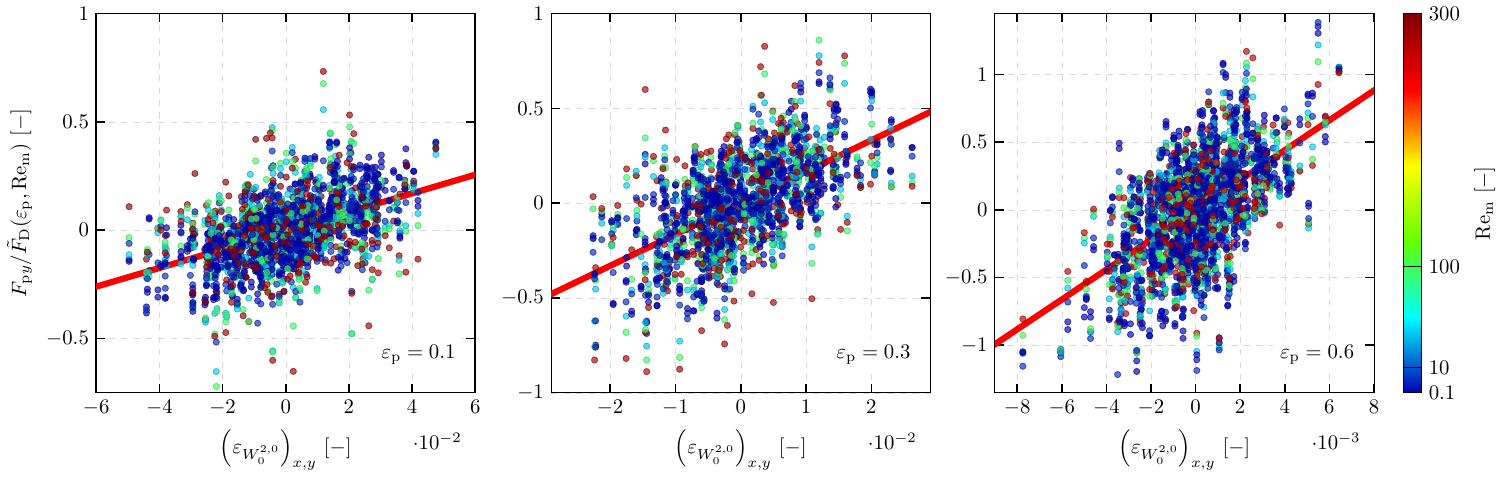}
\includegraphics[width=0.99\textwidth]{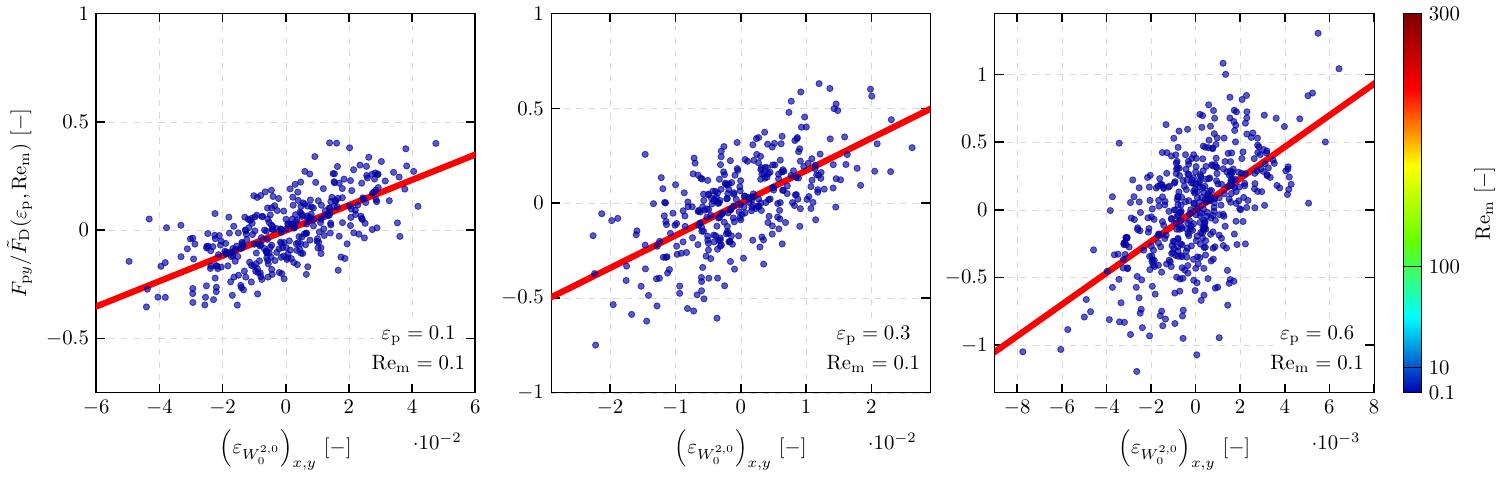}
\includegraphics[width=0.99\textwidth]{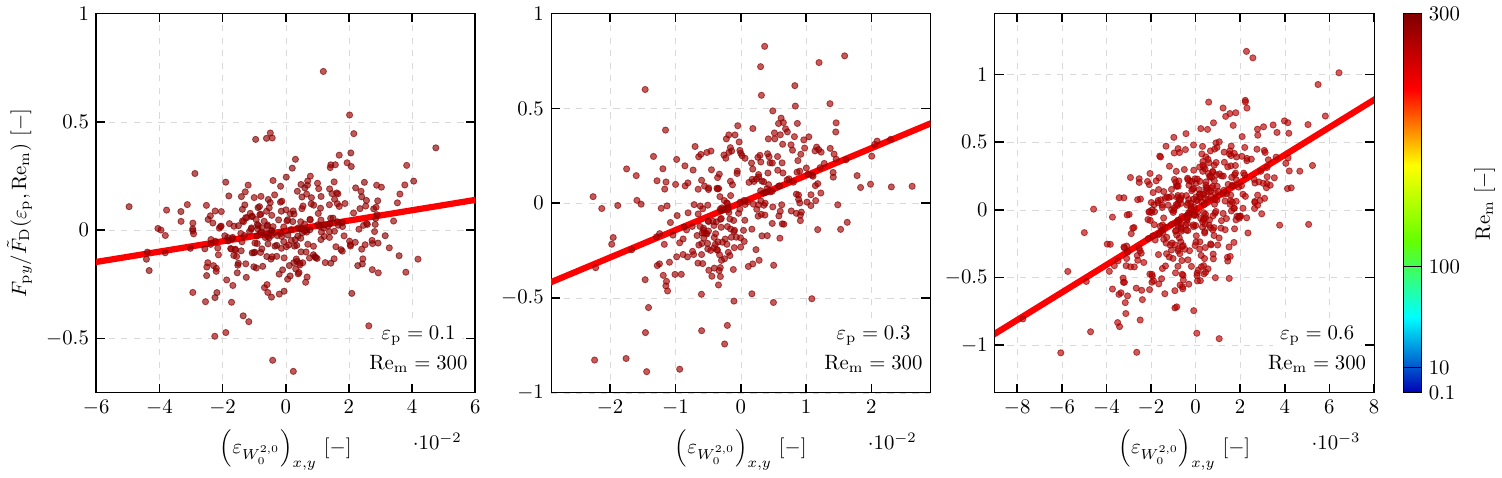}
\caption{Scatter plots of the relative lift force ($y$ component) of the individual particles as a function of the $x,y$ component of the Minkowski tensor $\tens{\varepsilon}_{W^{\textup{2,0}}_{\textup{0}}}$. The top row shows the relative lift force for all mean Reynolds numbers for 3 different mean solid volume fractions, the middle row shows the relative lift force for the mean Reynolds number of 0.1  for 3 different mean solid volume fractions, and the bottom row shows the relative lift force for the mean Reynolds number of 300  for 3 different mean solid volume fractions. The red line indicates the regression trend between the variables.}
\label{fig:ScatterPlotFy Voro-WTwoZeroZerotensor-01}
\end{figure}

Figure~\ref{fig:ScatterPlotFy Voro-WTwoZeroZerotensor-01} shows the relative lift force on the individual particles in the assembly as a function of the $x,y$ component of the tensor $\tens{\varepsilon}_{W^{\textup{2,0}}_{\textup{0}}}$; the top row showing the results for all mean Reynolds numbers, the middle row showing the results of the lowest mean Reynolds number, and the bottom row showing the results for the highest mean Reynolds number. The trends for all the configurations studied in this paper are indicated by the red line, and are quite strong; as the value of the $x,y$ component of the tensor $\tens{\varepsilon}_{W^{\textup{2,0}}_{\textup{0}}}$ increases, the magnitude of the relative lift on the individual particle in the assembly increases strongly. This holds true for all mean flow Reynolds numbers. This increase is stronger as the mean solid volume fraction increases.

\FloatBarrier

\begin{figure}[htbp!]
\centering
\includegraphics[width=0.99\textwidth]{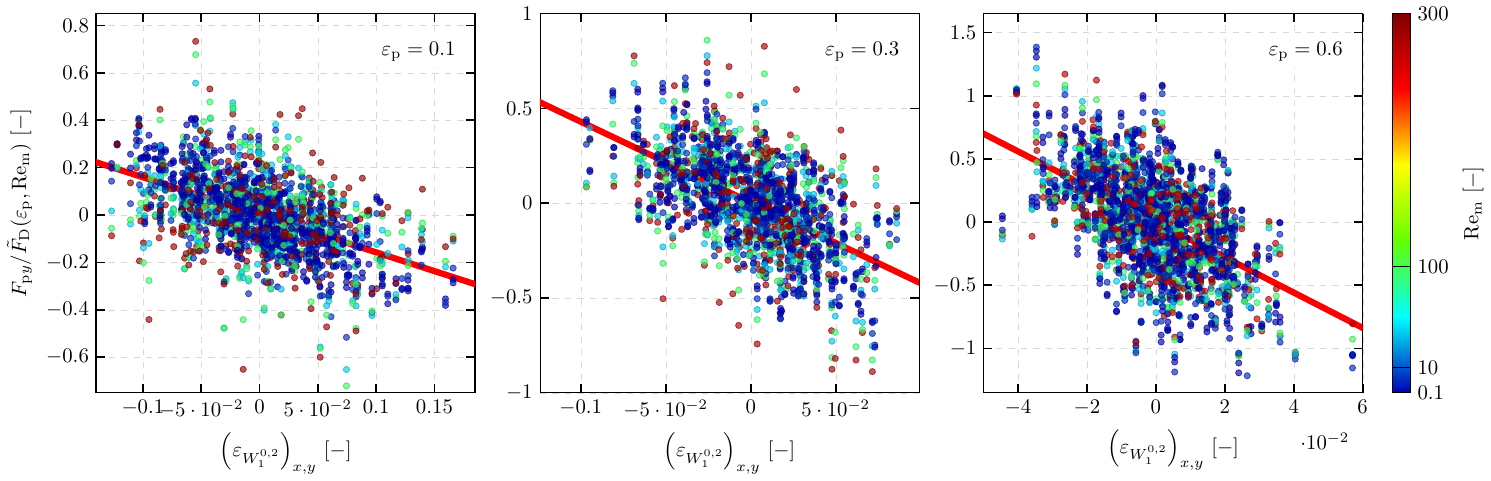}
\includegraphics[width=0.99\textwidth]{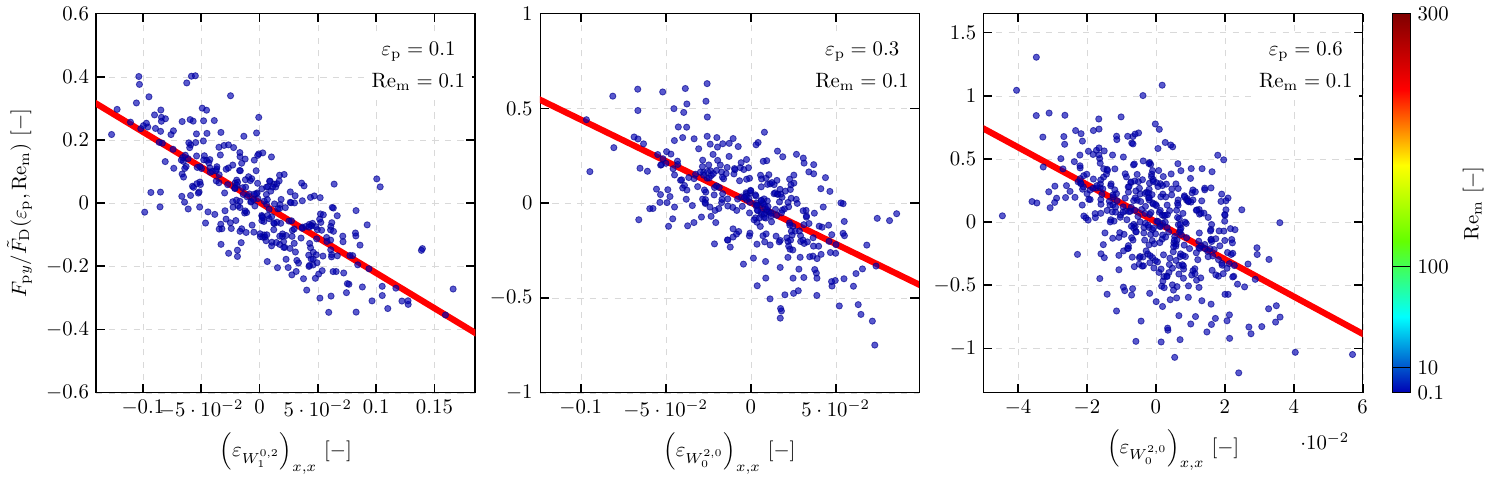}
\includegraphics[width=0.99\textwidth]{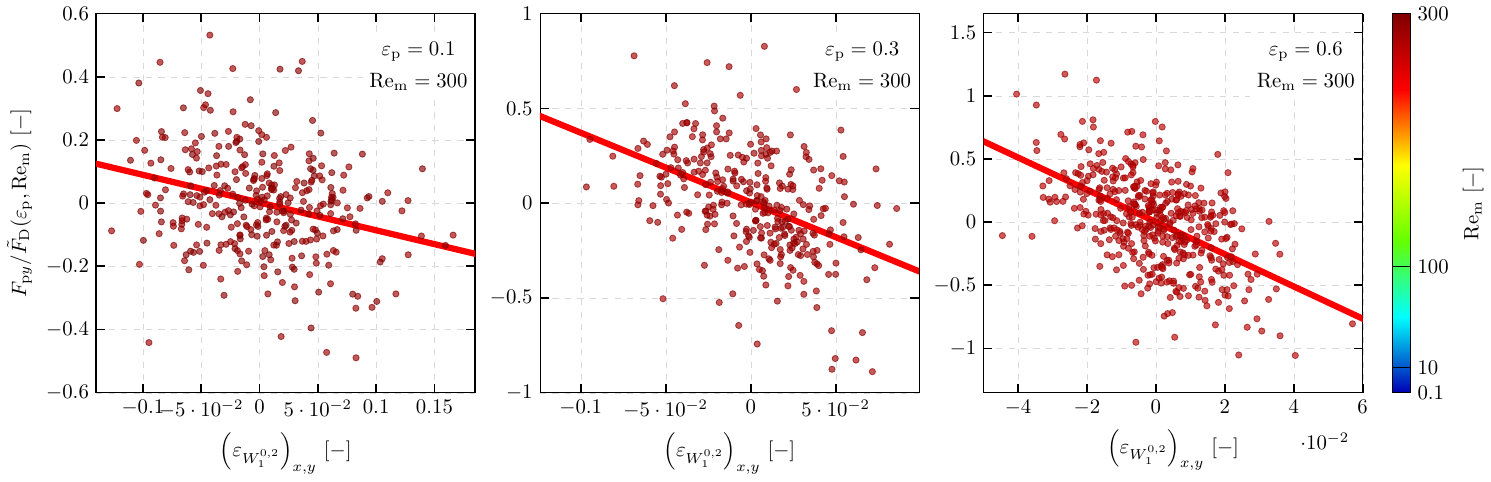}
\caption{Scatter plots of the relative lift force ($y$ component) of the individual particles as a function of the $x,y$ component of the Minkowski tensor $\tens{\varepsilon}_{W^{\textup{0,2}}_{\textup{1}}}$. The top row shows the relative lift force for all mean Reynolds numbers for 3 different mean solid volume fractions, the middle row shows the relative lift force for the mean Reynolds number of 0.1  for 3 different mean solid volume fractions, and the bottom row shows the relative lift force for the mean Reynolds number of 300  for 3 different mean solid volume fractions. The red line indicates the regression trend between the variables.}
\label{fig:ScatterPlotFy Voro-WZeroTwoOnetensor-01}
\end{figure}

An opposite trend is observed for the correlation between the relative lift force on the individual particles in the assembly as a function of the
$x,y$ component of the tensor $\tens{\varepsilon}_{W^{\textup{2,0}}_{\textup{0}}}$, as shown in figure~\ref{fig:ScatterPlotFy Voro-WZeroTwoOnetensor-01}.
As in the previous section, the top row showing the results for all mean Reynolds numbers, the middle row showing the results of the lowest mean Reynolds number, and the bottom row showing the results for the highest mean Reynolds number.
The trends for all the configurations studied in this paper are indicated by the red line, and are quite strong; as the value for the $x,y$ component of the tensor $\tens{\varepsilon}_{W^{\textup{2,0}}_{\textup{1}}}$ increases, the magnitude of the relative lift on the individual particle in the assembly decreases strongly. This holds to be true for all mean flow Reynolds numbers. This increase is stronger as the mean solid volume fraction increases.

\FloatBarrier

\subsection{Development of the microstructure-informed relative drag and lift force models}
\label{sec: Anisotropic results}

The model used to predict the forces acting on individual particles in the assembly is based on the mean drag force of the assembly, and a term describing the deviation of particle forces due to local anisotropy in the assembly compared to the mean drag force. This general expression is given in dimensionless form as
\begin{equation}
\vect{\tilde{F}}_{\mathrm{p}} = \tilde{F}_\mathrm{D}(\varepsilon_{\mathrm{p}}, \text{Re}_\textup{m})\left[\vect{e_{x}} + \vect{\tilde{F^{'}}}_{\mathrm{p}}\right]
\label{eq: General Model}
\end{equation}
where $\tilde{F}_\mathrm{D}(\varepsilon_{\mathrm{p}}, \text{Re}_\textup{m})$ is the dimensionless mean drag force (given by Eq.~\eqref{eq:MeanDragForceCorrelation}), and $\vect{\tilde{F^{'}}}_{\mathrm{p}}$ is the dimensionless term predicting the deviation from the mean force of the assembly.
In this work, we have chosen a configuration where the fluid flow is aligned with the $x$-direction, hence the direction $\vect{e_{x}}$ in the expression.
As a result of this choice, the correlation can be applied exclusively with Minkowski tensors corresponding to a fluid flow in the $x$-direction, which can be easily obtained by applying the rotation between the actual reference frame and the $x$-direction.

The correlations to predict the deviation of the force on the individual particles normalized by the mean drag force are given as functions of ``global'' quantities, \textit{i.e.} the mean flow Reynolds number and the solid volume fraction of the assembly, and the parameters describing the local microstructure based on the Minkowski tensors.
The general expression to predict the force deviation from the mean drag force of the assembly is given by
\begin{align}\label{eq: Model Predicting Fluctuation}
  \vect{\tilde{F^{'}}}_{\mathrm{p}}\left(\text{Re}_\textup{m},\varepsilon_{\mathrm{p}}, \tens{\varepsilon}, \vect{\varepsilon}, \varepsilon \right) = &
  {\tilde{F}}_{\mathrm{p}x}^{'}\left(\tens{\varepsilon} \cdot \vect{e_{x}} \cdot \vect{e_{x}},\vect{\varepsilon} \cdot \vect{e_{x}},\varepsilon\right) \cdot \vect{e_{x}} \nonumber \\
  + &
  {\tilde{F}}_{\mathrm{p}y}^{'}\left(\tens{\varepsilon} \cdot \vect{e_{y}} \cdot \vect{e_{y}},\vect{\varepsilon} \cdot \vect{e_{y}},\varepsilon\right) \cdot \vect{e_{y}} \\
  + &
  {\tilde{F}}_{\mathrm{p}z}^{'}\left(\tens{\varepsilon} \cdot \vect{e_{z}} \cdot \vect{e_{z}},\vect{\varepsilon} \cdot \vect{e_{z}},\varepsilon\right) \cdot \vect{e_{z}}\nonumber 
\end{align}
where the first term on the right-hand side is the relative deviation in the drag force, and the second and third terms on the right-hand side are the two lift force components.
As outlined in Section~\ref{sec: Lift force}, the same correlation is used to determine the values of the two lift force components, by alternating the Cartesian components of the expression.

To derive the expression to predict the deviation in the drag force and the two lift force components, symbolic regression is used to elucidate the complex relationship between the most important terms describing the microstructure and the forces on each of the particles. Several algorithms have been developed for symbolic regression~\citep{Udrescu2020,Biggio2021,Kaptanoglu2022}, and in this study we have adopted the PySR library~\citep{Cranmer2023}.
The regression consists of an internal search algorithm, which utilizes an evolve-simplify-optimize loop, and efficiently optimizes the form of several empirical expressions. After finding a number of suitable expressions,
we optimize the scalar constants within the newly-discovered empirical expressions. The expressions for the deviations in the drag force and the lift force on the individual particles in the assembly are described separately in the next sections. 

\subsubsection{Expression for predicting the deviation from the mean drag force}
\begin{figure}[htbp!]
\centering
\includegraphics[width=1.05\textwidth]{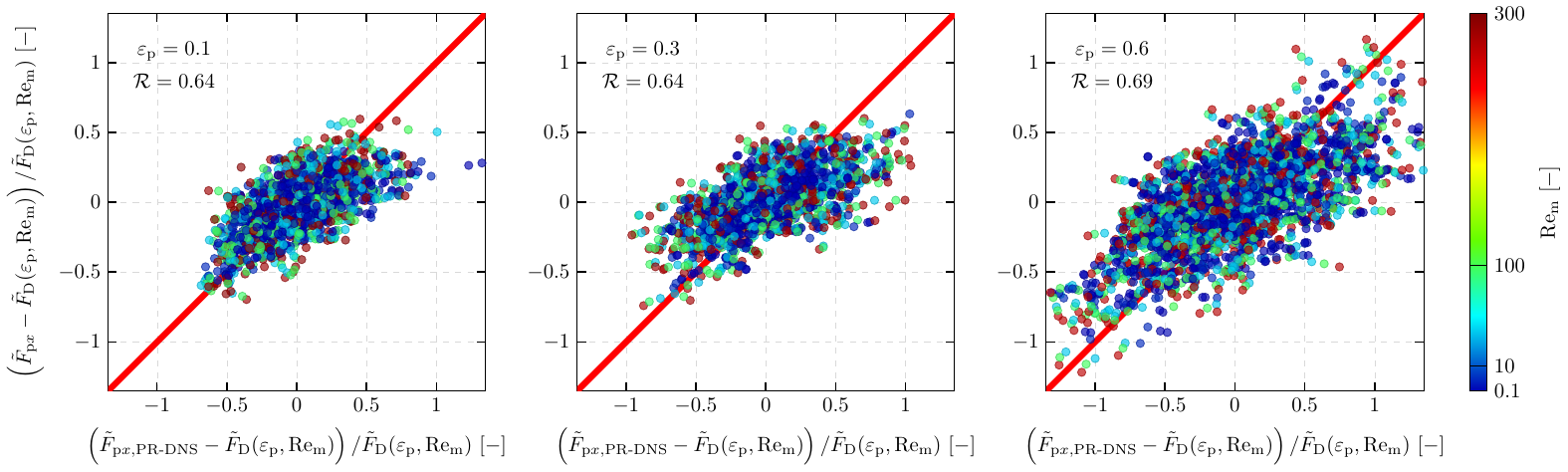}
\caption{Regression plots between the drag force deviation from the mean drag force term obtained from the PR-DNS, $\tilde{F}_{\mathrm{p}x}^{'}$, and the predicted values using the microstructure based deterministic model given in Eq.~\eqref{eq: Prediction Deviation drag force}. The Pearson correlation coefficient is given for three solid volume fractions, $\varepsilon_{\mathrm{p}} = 0.1, 0.3$ and $0.6$, for all flow regimes.}
\label{fig:ScatterPlotPrediction Drag Deviation term}
\end{figure}

The expression to predict the deviation in the drag force of individual particle normalized by the mean drag force of the assembly is given by
\begin{align}\label{eq: Prediction Deviation drag force}
\left[1 -0.841 \varepsilon_{\mathrm{p}}\right]^{-1} \tilde{F}_{\mathrm{p}x}^{'} = & \frac{\left[\varepsilon_{V} + \left(\varepsilon_{W^{\textup{0,2}}_{\textup{1}}}\right)_{xx}\left(8.563 - 0.787\text{ exp}^{\left(0.976^{\text{Re}_\textup{m}}\right)}\right)\right]}{\beta\left(\varepsilon_{W^{\textup{0,2}}_{\textup{1}}}\right)_{xx}\log{\left(\varepsilon_{\mathrm{p}}\right)}}\times\\
& \frac{\left[2.618\left(\varepsilon_{\mathrm{p}} - 0.335\right)\left( \beta \varepsilon_{\mathrm{p}}- \varepsilon_{V}\right) + \left(\left(\varepsilon_{W^{\textup{0,2}}_{\textup{1}}}\right)_{xx} - 0.335 \beta (0.824^{\text{Re}_\textup{m}} + 0.314)^{\left(\varepsilon_{W^{\textup{1,0}}_{\textup{0}}}\right)_{x}}\right)\right]}{\beta\left(\varepsilon_{W^{\textup{0,2}}_{\textup{1}}}\right)_{xx}\log{\left(\varepsilon_{\mathrm{p}}\right)}}\nonumber
\end{align}
with
$$
\beta = (1.067^{\varepsilon_{V}})^{18.523 -2.182 \times 10^{3}\left(\varepsilon_{W^{\textup{2,0}}_{\textup{1}}}\right)_{xx}}
$$
Figure~\ref{fig:ScatterPlotPrediction Drag Deviation term} shows the drag force on individual particle predicted by this proposed model compared to the PR-DNS data generated in this work, for three different solid volume fractions, 
$\varepsilon_{\mathrm{p}} = 0.1, 0.3$ and $0.6$ (first three figures), and for all solid volume fractions (last figure).  Also, the Pearson correlation coefficient is shown in the figures for the expressions, and lies above 0.67 for the whole dataset.

\subsubsection{Expression for predicting the lift force}
\begin{figure}[htbp!]
\centering
\includegraphics[width=1.05\textwidth]{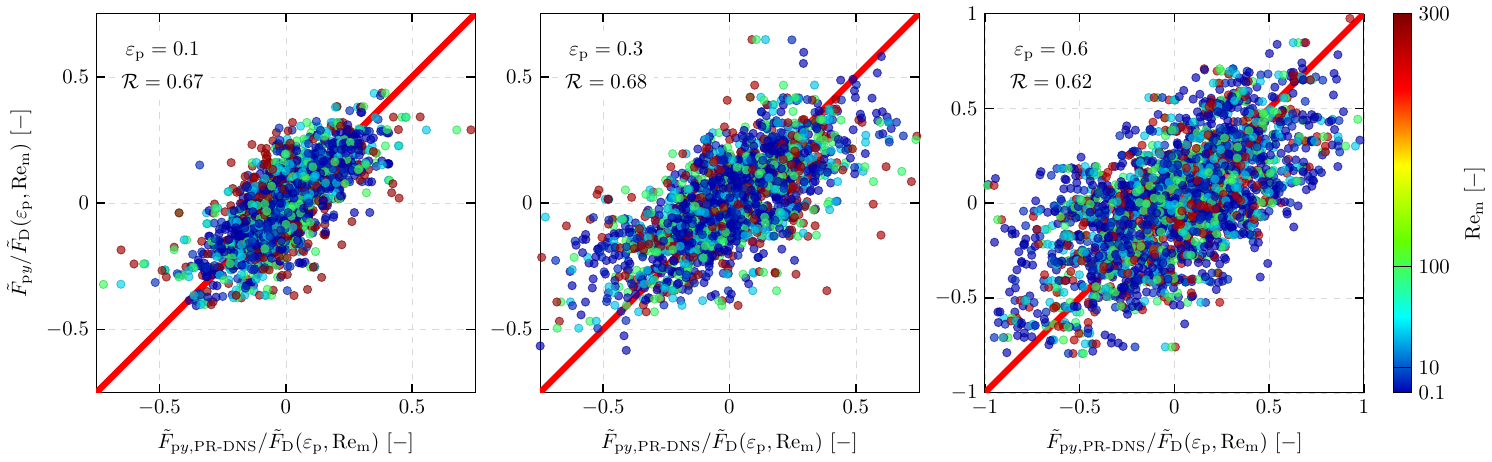}
\caption{Regression plots between the lift force deviation from the mean drag force term obtained from the PR-DNS, $\tilde{F}_{\mathrm{p}x}^{'}$, and the predicted values using the microstructure based deterministic model given in Eq.~\eqref{eq: Prediction Deviation drag force}. The Pearson correlation coefficient is given for three solid volume fractions, $\varepsilon_{\mathrm{p}} = 0.1, 0.3$ and $0.6$, for all flow regimes.}
\label{fig:ScatterPlotPrediction Lift Deviation term}
\end{figure}

The expression to predict the lift force on individual particles is given by
\begin{align}\label{eq: Prediction Deviation lift force}
  \frac{\left[\gamma + 27.341\right]}
 {\left[1 + 1.65 \varepsilon_{\mathrm{p}}\right]}\tilde{F}_{\mathrm{p}i}^{'} = & \left[2.878\left(-\left(\varepsilon_{W^{\textup{0,2}}_{\textup{1}}}\right)_{xi} + \left(\varepsilon_{W^{\textup{2,0}}_{\textup{1}}}\right)_{xi}\right)\left(\gamma + 26.183\right)\right.+\\
&\left.\left(\varepsilon_{\mathrm{p}}\left(\left(\varepsilon_{W^{\textup{1,0}}_{\textup{0}}}\right)_{y} + \left(\gamma + 26.063\right)\left(0.348\left(\varepsilon_{W^{\textup{0,2}}_{\textup{1}}}\right)_{xi} + 30.349\left(\varepsilon_{W^{\textup{2,0}}_{\textup{1}}}\right)_{xi} - 2\left(\varepsilon_{W^{\textup{2,0}}_{\textup{0}}}\right)_{xi}\right)\right) -\right.\right.  \nonumber\\
 &\left.\left.\mathcal{D}_{i} - \left(\varepsilon_{W^{\textup{2,0}}_{\textup{0}}}\right)_{xi}\right)\text{exp}^{\alpha}\right]\nonumber\,
\end{align}
with
\begin{equation}
\gamma = \text{exp}^{\left(\text{Re}_\textup{m}\varepsilon_{V}\right)^{-1}}\,
\end{equation}
and
\begin{equation}
\alpha = 3.867\text{exp}^{\left(0.0378\left(\varepsilon_{W^{\textup{1,0}}_{\textup{0}}}\right)_{i}\right)}\,
\end{equation}
In Eq.~\eqref{eq: Prediction Deviation lift force} $i$ = $y$ or $z$, and $j$ is $\neg i$, so $z$ or $y$, respectively. It should be noted, that \textit{two} lift forces are required, one in the $y$-direction ($i$ = $y$ and $j$ = $z$) and one in the $z$-direction ($i$ = $z$ and $j$ = $y$), compared to the direction of the fluid flow, which is assumed to be aligned with the $x$-direction.
Figure~\ref{fig:ScatterPlotPrediction Lift Deviation term} shows the $y$ component of the lift force on the individual particles predicted by the microstructure-based model compared to the PR-DNS data generated in this work, for the three different solid volume fractions, $\varepsilon_{\mathrm{p}} = 0.1, 0.3$ and $0.6$ (first three figures) and for all solid volume fractions (last figure). Also, the Pearson correlation coefficient is shown in the figures for the expression, and lies around 0.62 or higher for the whole dataset.

\subsection{Model verification}
\label{sec: Anisotropic results}
To assess the capability of the microstructure-based model to predict the forces on individual particles in an assembly of monodispersed fixed spheres, additional PR-DNS of the fluid flow past a random assembly of particles at various flow regime and solid volume as given in table~\ref{tab:Microstructure-based model RMS Error} are carried out. The particle assemblies are generated randomly and these have not been used to generate any of the models put forward in this work. This first two columns specify the mean solid volume fraction in the assembly and the mean particle Reynolds number of the flow through the assembly.
The third column of table~\ref{tab:Microstructure-based model RMS Error} shows the error in the average drag on the assembly compared to the average drag model, given by Eq.~\ref{eq:MeanDragForceCorrelation}. This error is typically a few percent for all studied cases.

\begin{figure}[htbp!]
\centering
\includegraphics[width=1.\textwidth]{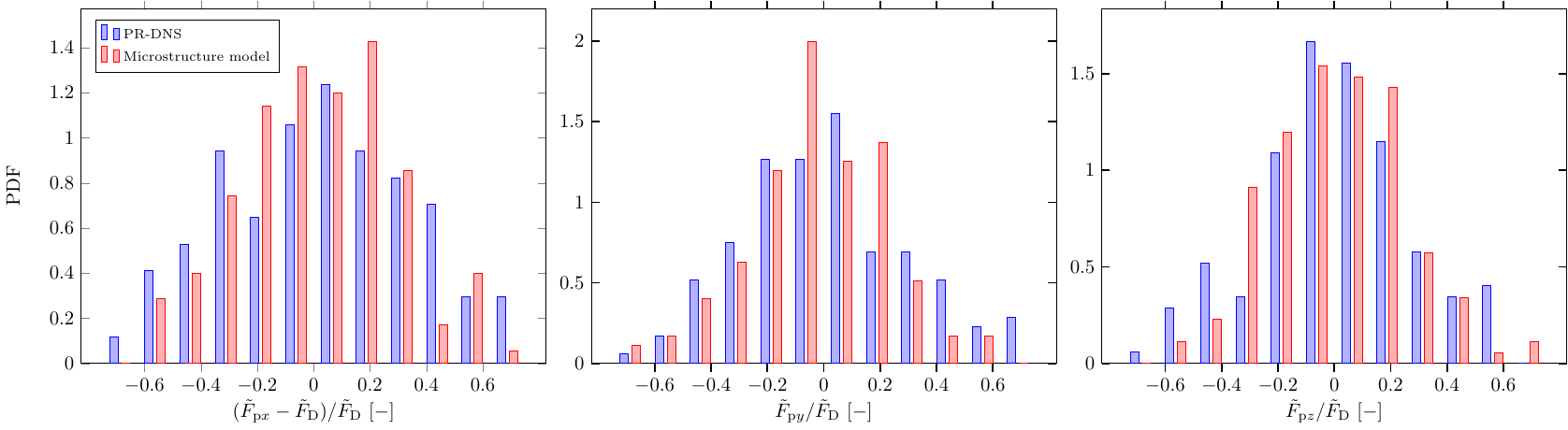}
\caption{Probability density function of the normalized dimensionless drag force ($\tilde{F}_{\mathrm{p}x}$) and the two lift force components ($\tilde{F}_{\mathrm{p}y}$ and $\tilde{F}_{\mathrm{p}z}$), from left to right, for the flow configuration $\left(\varepsilon_{\mathrm{p}},\mathrm{Re_m}\right) = \left<0.45,135.96\right>$.}
\label{fig:Histogram Comparison Model vs PRDNS}
\end{figure}

To analyze the microstructure models put forward in this work, \textit{i.e.} Eqs.~\ref{eq: Model Predicting Fluctuation} and ~\ref{eq: Prediction Deviation lift force}, we plot the probability density function of the dimensionless drag force ($\tilde{F}_{\mathrm{p}x}$) and the two lift force components ($\tilde{F}_{\mathrm{p}y}$ and $\tilde{F}_{\mathrm{p}z}$) for the flow configuration $\left(\varepsilon_{\mathrm{p}},\mathrm{Re_m}\right) = \left<0.45,135.96\right>$, and report the results for the PR-DNS and microstructure models in figure~\ref{fig:Histogram Comparison Model vs PRDNS}.
For all force components the results demonstrate the ability to capture the mean tendency and shape of the distribution resembling a normal distribution. However, the microstructure models slightly underestimate the variability or spread of the data as reflected by the underestimation of the standard deviation.

To evaluate the difference between the hydrodynamic forces of the PR-DNS results and the microstructure models put forward in this work, the following expression is used: 
\begin{equation}
Er_i = \frac{1}{N_\mathrm{p}} \sum_{i=1}^{N_\mathrm{p}} \cfrac{\left(\tilde{F}_{\mathrm{p}i,\mathrm{PR-DNS}} - \tilde{F}_{\mathrm{p}i,\text{model}}\right)^{2}}{\sigma\left(\tilde{F}_{\mathrm{p}i,\mathrm{PR-DNS}}\right)}
\end{equation}
The fourth, fifth and sixth columns of table~\ref{tab:Microstructure-based model RMS Error} indicate the component-wise RMS error which is made if the averaged drag model is used, and the 7th, 8th and 9th columns show the error with the newly proposed microstructure informed drag model.
For all flow configurations, the error made in the prediction of the hydrodynamic forces on the individual particles as compared to the PR-DNS is reduced when using the microstructure based models. However, the accuracy of the model varies per configuration.
For instance, the RMS error in the drag prediction is decreased by a factor of two for the flow configuration $\left(\varepsilon_{\mathrm{p}},\mathrm{Re_m}\right) = \left(0.35,65.86\right)$ when compared to the averaged drag force model. However, there are also configurations where the improvement is significantly less.

\begin{table}[h]
\centering
    \begin{tabular}{|l|l|c|ccc|ccc|} \hline
       & 
       &  average error of
       & \multicolumn{3}{c|}{RMS errors assuming averaged drag}
       & \multicolumn{3}{c|}{RMS errors with microstructure drag}\\
       $\varepsilon_{\mathrm{p}}$ & $\mathrm{Re_m}$ 
       & averaged drag model
       & drag & lift$_y$ & lift$_z$ 
       & drag & lift$_y$ & lift$_z$ \\
    \hline
    \hline
    0.55 & 4.42& 4.30 \% & 33.30 \% & 26.74 \% &26.85\%
    & 22.96\% & 17.89\%  & 16.69\%  \\
    0.55 & 175.51 & 4.67\% & 37.17 \% & 23.89\% & 23.21\%
    & 24.20 \% & 21.19\% & 18.56\% \\
    0.45 & 135.96  & 3.14 \% & 36.22\% & 30.05 \% & 27.68\%
    & 20.76 \%& 15.77 \% & 16.91\% \\
    0.35 & 65.86 &  4.11 \% & 37.09\% & 26.14 \% & 26.07\%
    & 18.11 \% & 15.22\%  & 14.59\% \\
    \hline
    \end{tabular}
    \caption{Results of four different test cases of particle assemblies, showing the error in the average drag model (3rd column), the RMS error in the case of assuming an averaged drag model (columns 4 to 6), and the RMS error in the case of using a microstructure informed drag model (columns 7 to 9).}
    \label{tab:Microstructure-based model RMS Error}
\end{table}

\section{Model implementation in the CFD/DEM framework}
\label{sec: ModelImplementation}
\begin{figure}[htbp!]
\centering
\includegraphics[width=0.65\textwidth]{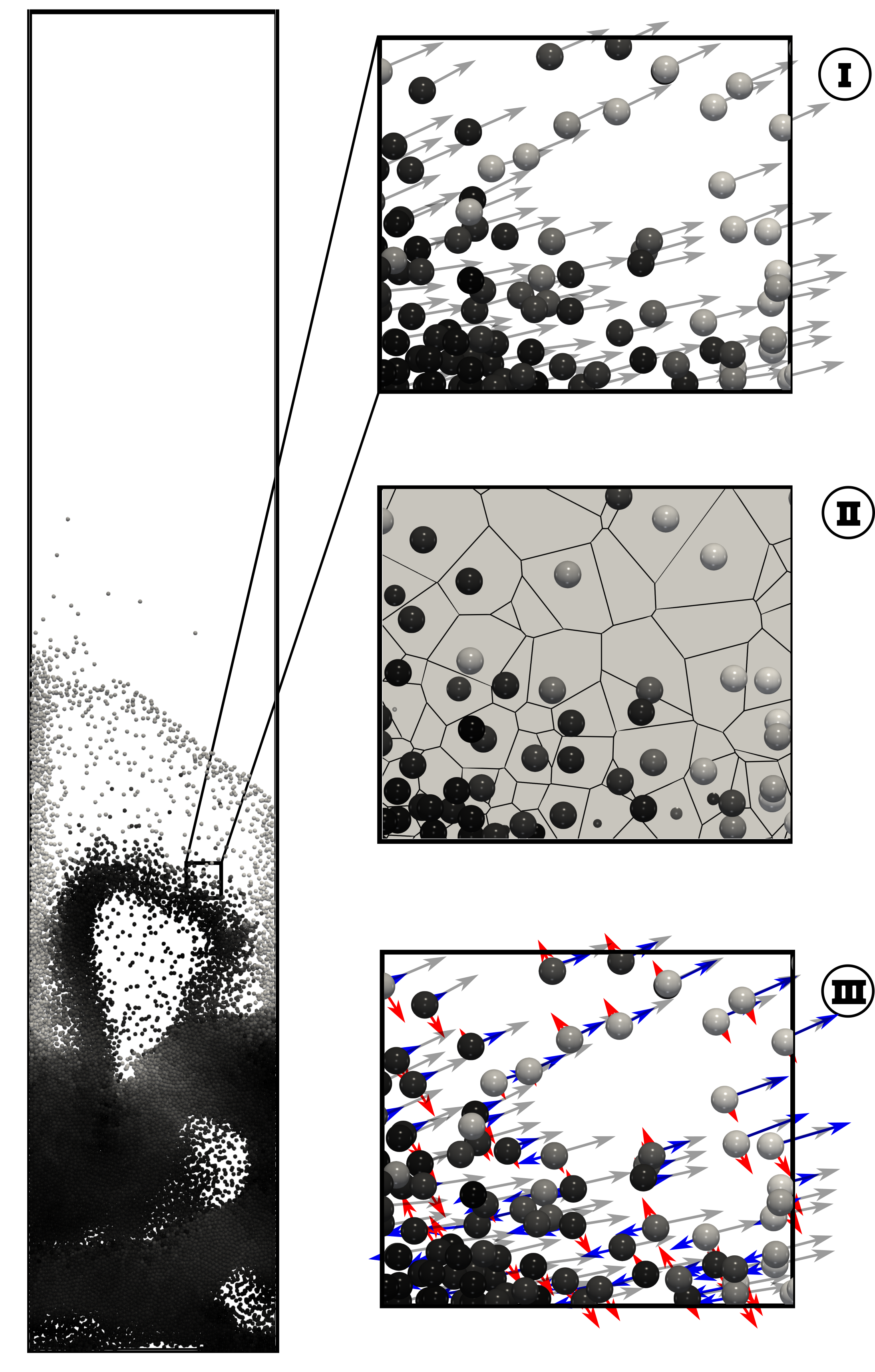}
\caption{Implementation and application of the microstructure-based hydrodynamic force model in the CFD/DEM framework. $\textbf{I}$ shows a cross-section of a computational cell with the particles and their `averaged' drag force (as grey arrows), $\textbf{II}$ shows the Voronoi tessellation of the particle assembly, and $\textbf{III}$ shows the drag force deviation (in blue) and the lift force (in red) on the individual particles based on the novel expressions developed in this work.}
\label{fig:CFD-DEM Implementation}
\end{figure}

In this work we have put forward a microstructure based hydrodynamic force model, to determine the force on the individual particles in an assembly, with solid volume fractions up to 0.6 and mean particle Reynolds numbers up to 300.
This microstructure-based hydrodynamic force model has a large potential to increase the accuracy of Eulerian-Lagrangian or CFD/DEM simulation frameworks. 
Moreover, implementing this model is straightforward in any existing Eulerian-Lagrangian or CFD/DEM framework, as it does not require any modifications to the flow solver or particle transport; the modifications are limited to the computation of the forces on each individual particle. The procedure is sketched in
Figure~\ref{fig:CFD-DEM Implementation}, illustrating a CFD/DEM simulation of fluidization with spherical particles. For each time-step, the solid volume fraction and mean flow Reynolds number are computed for group of particles, typically the particles in a Eulerian fluid cell to determine the mean drag force on the particles in that assembly using Equation~\ref{eq:MeanDragForceCorrelation}. Subsequently, the entire particle assembly undergoes a Voronoi tessellation, and the relevant Minkowski tensors are computed in the reference frame of the mean local fluid velocity; \textit{i.e.} aligning the mean relative flow direction with the $x$-direction. Based on the properties of the local Voronoi tessellation, three additional forces are computed, the relative drag force, see Equation~\ref{eq: Prediction Deviation drag force}, and the two lift forces using Equation~\ref{eq: Prediction Deviation lift force}. By adding these three forces to the forces on the individual particle, Newton's second law can be applied to update the location of the particles in time.
The implementation of the microstructure-based model to an existing CFD/DEM framework is summarized in Algorithm~\ref{alg:CFD-DEM Implementation}, with the additional steps indicated by a star symbol ($\star$).

\RestyleAlgo{ruled}
\SetKwComment{Comment}{/* }{ */}

\begin{algorithm}[hbt!]
\caption{Implementation of the microstructure-based correlations to predict individual particle forces in Euler/Lagrange, or CFD-DEM, approach. Additional steps required for the novel model are indicated with a star symbol ($^{\star}$).}\label{alg:CFD-DEM Implementation}
\For{each time level t}{
    \For{each Eulerian fluid cell $i$}{
    Update list of particles per fluid cell.\\
    Compute Eulerian fluid cell solid volume fraction $\varepsilon_{\mathrm{p},i}$.\\
    }
    Perform Voronoi tessellation.$^{\star}$\\
    \For{each Lagrangian paricle $k$,}{
    Interpolate the gas velocity to the center of particle $k$.\\
    Compute the mean drag force $\tilde{F}_{\mathrm{p,D,k}}$ (Eq.~\eqref
    {eq:MeanDragForceCorrelation}).\\
    Rotate the Voronoi cell to main local fluid velocity reference frame.$^{\star}$\\
    Compute the Voronoi tensors.$^{\star}$\\
    Compute the particle force deviation $\vect{\tilde{F}}^{'}_{p,k}$ (Eq. 30-31).$^{\star}$\\
    $\Rightarrow$ \textit{Compute additional external forces based on reduced order model.}\\
    Move particle $k$ using Verlet integration scheme to solve the particle transport equation.\\
    \If{Two-way coupling}{
    Update momentum contribution of particle $k$ to the fluid mesh.\\
    }
    }
    Solve Eulerian fluid governing equations.\\
}
\end{algorithm}

\section{Conclusion}
\label{sec: Conclusion}
In this work, we have put forward hydrodynamic force models to accurately describe the interaction of a flow with the individual particles in an assembly, when only an averaged description of the flow is available. In total, there are three new hydrodynamic force models put forward in this work. These hydrodynamic force models are based upon a large number of PR-DNS, carried out in this work. The first hydrodynamic force model predicts the \textit{average drag} on an assembly of particles. Although the form of this model is similar to the expression of~\citet{Tenneti2011}, we have included data from PR-DNS with a higher solid volume fraction, so it is better suitable for dense particle-laden flows.

The second hydrodynamic force model put forward in this work predicts the \textit{deviation of the drag} force for the individual particles in the assembly. This model is based upon the characterization of the local microstructure of the particles in the assembly, obtained through the Minkowski tensors of the Voronoi tessellation. Finally, the third hydrodynamic force model predicts the additional \textit{lift forces} on the individual particles in the assembly. Although a lift force on the particle assembly as a whole are typically ignored, this force may be important to predict the motion of the individual particles. These three hydrodynamic force models are based upon the best fits from all the PR-DNS data obtained in this work using symbolic regression.

In this paper we also describe how to implement the model in an existing Eulerian-Lagrangian or CFD/DEM framework, and show an improvement in the prediction of the hydrodynamic forces compared to use only an averaged drag force model, as is usually done.
By analyzing multiple flow cases, we show that, for these cases, the reduction of the error in the prediction of the hydrodynamic forces ranges from a few percent to a factor of two.
In addition, the microstructure models have the ability to capture the tendency and shape of the distribution of forces over an assembly of particles.

In conclusion, our study presents novel hydrodynamic force models that improve the prediction of particle-fluid interactions on individual particles based on the local anisotropy of the particle assembly in scenarios where detailed flow information is lacking. These models, when appropriately applied, exhibit noticeable improvement over conventional averaged drag force models, thus establishing a pathway towards more accurate simulations of particle-laden flows in diverse engineering domains.

\section*{Data Availability Statement}
The data that support the findings of this study are reproducible
and files to regenerate the correlations are openly available in the repository with DOI
10.5281/zenodo.8282784, available at \href{https://doi.org/10.5281/zenodo.8282784}{https://doi.org/10.5281/zenodo.8282784}

\section*{Acknowledgements}
\noindent 
This work was funded by the Deutsche Forschungsgemeinschaft (DFG, German Research Foundation) - Project-ID 422037413 - TRR 287.

\bibliographystyle{model1-num-names}

\end{document}